%% file: main.tex
\documentclass[conference]{IEEEtran}

\usepackage{comment}

\usepackage{graphicx}
\usepackage{booktabs} 
\usepackage{array}    
\usepackage{tabularx}
\usepackage[linesnumbered, ruled, vlined]{algorithm2e}
\usepackage{amsmath,amssymb,amsfonts}
\usepackage{multirow}

\usepackage{newtxtext, newtxmath}  
\usepackage{xcolor}
\usepackage{subcaption}
\usepackage{makecell}
\usepackage{pgf-pie}
\usetikzlibrary{patterns}
\usepackage{ragged2e}
\usepackage{cite}
\usepackage{url}
\usepackage[hidelinks]{hyperref}

\usepackage{marvosym}
\usepackage{tcolorbox}
\usepackage{enumitem}
\definecolor{titlebg}{RGB}{200, 200, 200}        
\definecolor{boxbg}{RGB}{250, 250, 250}       
\definecolor{boxborder}{RGB}{200, 200, 200}      

\newcommand{\lbz}[1]{{[\textcolor{red}{LBZ:} \textcolor{cyan}{#1}]}}

\ifCLASSINFOpdf
\else
\fi
\begin{document}

\bstctlcite{IEEEexample:BSTcontrol} 

%
\title{The Colossus with Feet of Clay: Debunking Encrypted Traffic Classifiers under PQC Evolution}


\author{
\IEEEauthorblockN{
Bingzhen Li\textsuperscript{1}, 
Lingjia Meng\textsuperscript{2,\textsuperscript{\Letter}}, 
Runhan Song\textsuperscript{2},
Chuanzhou Pan\textsuperscript{2},
Tongjun Pu\textsuperscript{3},
Ziqiang Ma\textsuperscript{3}, \\
Yupeng Jiang\textsuperscript{1},
Lei Cui\textsuperscript{2},
Zhiyu Hao\textsuperscript{2}
}

\IEEEauthorblockA{\textsuperscript{1}\textit{School of Cyber Science and Technology, Beihang University}, Beijing, China \\
\textsuperscript{2}\textit{Zhongguancun Laboratory}, Beijing, China  \\
\textsuperscript{3}\textit{School of Information Engineering, Ningxia University}, Ningxia, China  \\
}
}
\maketitle

\begin{abstract}

Encrypted traffic classifiers often achieve high accuracy under matched training and testing conditions, implicitly assuming that deployment traffic follows the training distribution. TLS migration toward post-quantum cryptography (PQC) challenges this assumption because hybrid key establishment can reshape observable traffic without changing application labels. We frame this change as PQC-induced protocol drift and study its effects through closed-world HTTPS website fingerprinting using the deployed TLS~1.3 Hybrid-PQC group \texttt{\detokenize{X25519MLKEM768}}. We build a controlled, PQC-aware benchmark pairing Traditional (Non-PQC) and Hybrid-PQC traffic, then evaluate five representative classifiers and side-channel representations under matched-domain, cross-domain, and deployment-ratio settings. Collectively, the experiments show that PQC evolution does not remove learnable website information. Instead, it changes how that information appears in traffic, causing classifiers and feature combinations that perform well in-domain to lose reliability across cryptographic domains. By exposing the fragility of matched-domain evaluation, we offer strategic guidance, identify cross-domain robustness as a research priority, and recommend protocol-aware practices for dependable real-world encrypted traffic classification.
The code is available at \href{http://anonymous.4open.science/r/PQ-WF-Eval}{\nolinkurl{http://anonymous.4open.science/r/PQ-WF-Eval}}.

\end{abstract}



%
\IEEEpeerreviewmaketitle

\input{Tex/1-Introduction}
\input{Tex/2-Background}
\input{Tex/3-Analysis}
\input{Tex/4-Dataset}

\input{Tex/5-Evaluation}
\input{Tex/6-Discussion}
\input{Tex/7-Conclusion}






%

\bibliographystyle{IEEEtran}
\bibliography{mybib}

\end{document}

%% file: Tex/1-Introduction.tex
\section{Introduction}
\label{sec:intro}

Encrypted traffic classifiers are typically evaluated with training and test traffic drawn from the same distribution, implicitly assuming that this distribution remains representative after deployment~\cite{SoKDecodingEnigma2025a,zhaoSweetDangerSugar2025,akbariOneTaskRule2025}. One representative security task in this area is closed-world HTTPS website fingerprinting (WF)~\cite{chengHOLMESWATSON2025,cheng2026star}, in which a passive adversary infers the visited website without decrypting TLS payloads. The adversary does so by exploiting packet lengths, directions, timing, and byte-level packet representations~\cite{cumulWebsiteFingerprintingInternet2016,DeepFingerprintingUndermining2018a,TikTokUtilityPacket2020,ETBERTContextualizedDatagram2022a,YatcAnotherTrafficClassifier2023a}. Under matched training and testing conditions, prior WF methods have reported high classification accuracy. Such results, however, do not establish whether classifiers remain reliable as the underlying cryptographic protocol stack evolves.

The deployment of post-quantum cryptography (PQC) provides a concrete instance of this evolution. TLS migration currently centers on hybrid key establishment, which combines a traditional key exchange with a PQC key-encapsulation mechanism (KEM)~\cite{nistFIPS203MLKEM2024,alagicRecommendationsKeyEncapsulationMechanisms2025}. We focus on the deployed TLS~1.3 Hybrid-PQC group \texttt{\detokenize{X25519MLKEM768}}. Compared with the Traditional (Non-PQC) X25519, this group introduces substantially larger key shares and expands the corresponding handshake messages. These additional bytes can alter TCP segmentation, packet positions, and timing~\cite{RFC9954HybridKeyExchange2026,rfc9846TLS13,montenegroPerformancePQTLS2026}. Thus, although the website label is unchanged, the key-establishment mechanism can reshape the observable traffic trace used for classification.

Distribution drift arises when the traffic distribution observed during deployment differs from that represented in the training data. Within this broader category, we define \textit{protocol drift} as shifts in traffic distribution resulting from the evolution or deployment of protocol mechanisms. Unlike temporal, network, or application drift, protocol drift alters the traffic-generation process under otherwise fixed experimental conditions~\cite{rfc9846TLS13,rfc9000QUIC2021,alanClientDiversityHTTPS2019,chen2025drift}. The transition from Traditional (Non-PQC) X25519 to Hybrid-PQC \texttt{\detokenize{X25519MLKEM768}} key establishment provides a concrete instance of \textit{PQC-induced protocol drift}. We examine its effects through closed-world HTTPS website fingerprinting. Accordingly, this study investigates how the transition to PQC in TLS affects the reliability of encrypted traffic classifiers.

This transition presents two closely related challenges for evaluating encrypted traffic classifiers.
\begin{itemize}
\item \textbf{Representation shifts despite stable task semantics}. The larger key shares introduced by PQC key establishment increase TLS handshake message sizes and may alter packet-length sequences, burst boundaries, TCP segmentation, and byte-level alignment~\cite{nistFIPS203MLKEM2024,montenegroPerformancePQTLS2026,nistFIPS204MLDSA2024}. These changes reshape classifier inputs even though the website label and application semantics remain unchanged. Classifiers may consequently learn configuration-specific correlations that fail to transfer across cryptographic domains.
\item \textbf{Deployment heterogeneity and cross-domain mismatch}. The Internet is unlikely to transition directly from Traditional key establishment to a single PQC configuration. Traditional, Hybrid-PQC, and future Pure-PQC mechanisms may therefore coexist throughout an extended migration period~\cite{akbariOneTaskRule2025,RFC9954HybridKeyExchange2026}. Once deployed, a classifier trained in one cryptographic domain may encounter traffic from a mixture of configurations, some absent from its training data. Matched-domain accuracy alone cannot characterize this mismatch or its sensitivity to the Hybrid-PQC deployment ratio.
\end{itemize}

We examine these effects using a controlled, PQC-aware encrypted traffic benchmark. First, we measure the deployment of TLS key-establishment mechanisms across 2,060 Tranco-ranked websites and select the 200 highest-ranked websites supporting \texttt{\detokenize{X25519MLKEM768}} as collection candidates~\cite{pochatTrancoResearchOrientedTop,rfc9794PQTTerminology2025}. After validation, 195 websites remain in the shared label set. For each retained website and cryptographic configuration, we collect 120 complete visits, which form the \textsf{Non-PQC Website Dataset} and \textsf{PQC Website Dataset}, respectively. Four Mixed test sets, together with the two endpoint test sets, span six Hybrid-PQC deployment ratios from 0\% to 100\%. Under a common closed-world WF task with visit-level partitions, we evaluate \textsf{CUMUL}, \textsf{DF}, \textsf{Tik-Tok}, \textsf{ET-BERT}, and \textsf{YaTC} while retaining their native inputs and preprocessing pipelines~\cite{cumulWebsiteFingerprintingInternet2016,DeepFingerprintingUndermining2018a,TikTokUtilityPacket2020,ETBERTContextualizedDatagram2022a,YatcAnotherTrafficClassifier2023a}. The evaluation examines matched-domain learnability, bidirectional cross-domain transfer, sensitivity to Hybrid-PQC deployment ratios, and the contributions of side-channel features.

Matched-domain results show that website identity remains learnable in either cryptographic domain. Each classifier achieves comparable performance in the two matched-domain settings. For example, \textsf{ET-BERT} achieves accuracies of $0.9316$ and $0.9332$, while \textsf{YaTC} obtains $1.0000$ and $0.9991$. This learnability, however, does not ensure cross-domain reliability. When models trained on Non-PQC traffic are applied to Hybrid-PQC traffic, the accuracies of \textsf{CUMUL}, \textsf{ET-BERT}, and \textsf{YaTC} drop from $0.7731$, $0.9316$, and $1.0000$ to $0.1919$, $0.6255$, and $0.3582$, respectively. We next examine how this mismatch changes during migration. As the Hybrid-PQC deployment ratio increases, performance declines monotonically for four classifiers, whereas \textsf{YaTC} follows a non-monotonic trajectory; none maintains its Non-PQC baseline across the full range.
Finally, controlled comparisons of side-channel representations show that packet length is the most discriminative individual feature, although its advantage diminishes as the Hybrid-PQC deployment ratio increases. Direction and timing provide limited additional benefit. Overall, PQC migration creates classifier mismatch without eliminating learnable website information.

The main contributions are as follows.
\begin{itemize}
\item \textbf{PQC-aware traffic benchmark for isolating protocol drift}. To the best of our knowledge, we introduce the first encrypted traffic benchmark designed to isolate distribution shifts induced by PQC key establishment. Its paired Non-PQC and Hybrid-PQC datasets and controlled Mixed test sets support evaluation across cryptographic domains and deployment ratios while holding the label space and visit-level partitions fixed.
\item \textbf{Systematic evaluation of cross-domain classifier reliability under PQC-induced protocol drift}. We assess five representative classifiers spanning aggregate statistics, packet sequences, timing-aware inputs, and pre-trained byte-level representations. Contrasting matched-domain and cross-domain performance separates domain-specific learnability from transfer reliability. 
\item \textbf{Controlled analysis of side-channel feature transferability}. Using \textsf{DF} as a fixed architecture, we isolate the individual and combined contributions of packet direction, length, and timing. Packet length remains the strongest individual feature, yet no evaluated feature composition prevents cross-domain degradation. These findings offer practical guidance for feature engineering in traffic analysis as cryptographic protocols evolve.
\end{itemize}

The remainder of this paper is organized as follows. Section~\ref{sec:back} reviews background and related work. Section~\ref{sec:analy} analyzes PQC-induced changes in TLS~1.3 traffic and formulates the corresponding hypotheses. Section~\ref{sec:dataset} describes the construction of the PQC-aware benchmark. Section~\ref{sec:evalu} evaluates representative classifiers under PQC-induced protocol drift, Section~\ref{sec:dis} discusses the implications, and Section~\ref{sec:conclusion} concludes the paper.

%% file: Tex/2-Background.tex
\section{Background and Related Work}
\label{sec:back}

\subsection{Post-Quantum Key Establishment and Deployment in TLS}

The security of widely deployed public-key mechanisms, including RSA and elliptic-curve cryptography, depends on the hardness of integer factorization or discrete logarithms~\cite{rivestMethodObtainingDigital1978,koblitzEllipticCurveCryptosystems1987}. Shor's algorithm solves both problems in polynomial time on a sufficiently capable quantum computer~\cite{shorAlgorithmsQuantumComputation1994}. A cryptographically relevant quantum computer does not yet exist, but migration is already time-sensitive. An adversary can record long-lived encrypted traffic today and decrypt it after such a computer becomes available, a threat commonly described as ``harvest now, decrypt later"~\cite{rfc9794PQTTerminology2025}.

NIST responded to this threat through its Post-Quantum Cryptography standardization project. In August 2024, NIST finalized ML-KEM in FIPS~203, ML-DSA in FIPS~204, and SLH-DSA in FIPS~205~\cite{nistFIPS203MLKEM2024,nistFIPS204MLDSA2024,nistFIPS205SLHDSA2024}. ML-DSA and SLH-DSA provide digital signatures, whereas ML-KEM provides the key-establishment primitive relevant to this study. ML-KEM is derived from CRYSTALS-Kyber~\cite{bosCRYSTALSKyberCCASecure2018} and offers three parameter sets: ML-KEM-512, ML-KEM-768, and ML-KEM-1024.

A key-encapsulation mechanism comprises three algorithms: key generation (\textsf{KeyGen}), encapsulation (\textsf{Encaps}), and decapsulation (\textsf{Decaps}). Together, these algorithms allow two endpoints to establish a shared secret over a public channel~\cite{nistFIPS203MLKEM2024,alagicRecommendationsKeyEncapsulationMechanisms2025}. TLS~1.3 can supply that secret to its existing key schedule without changing the subsequent symmetric record-protection mechanisms. ML-KEM can therefore replace a traditional Diffie--Hellman exchange or be combined with one during migration.

Current web deployment is increasingly centered on hybrid key establishment~\cite{cloudflarePostQuantumInternet2025}. A post-quantum/traditional hybrid combines a post-quantum mechanism with a traditional key exchange~\cite{RFC9954HybridKeyExchange2026,rfc9794PQTTerminology2025}. Its combiner aims to preserve security if at least one component remains unbroken, subject to the stated security assumptions. The TLS hybrid group studied in this paper, \texttt{\detokenize{X25519MLKEM768}}, combines X25519 with ML-KEM-768~\cite{kwiatkowskiPostquantumHybridECDHEMLKEM2026}. Recent browser and TLS-stack deployments have enabled this group by default, making it a practical target for studying protocol evolution rather than a hypothetical configuration~\cite{cloudflarePostQuantumInternet2025}.

Throughout this paper, we distinguish three key-establishment settings to maintain a precise experimental scope. \textit{Traditional (Non-PQC)} key establishment uses X25519 without a post-quantum component, providing the baseline for our paired collections. \textit{Hybrid-PQC} combines X25519 with ML-KEM-768 through \texttt{\detokenize{X25519MLKEM768}} and defines the empirical PQC domain evaluated in this work. \textit{Pure-PQC} uses standalone ML-KEM key establishment and remains an active IETF specification effort~\cite{pureMLKEMPostQuantumKey2026a}.\footnote{We include \textit{Pure-PQC} only as a protocol-level reference in Section~\ref{sec:analy}; we neither collect \textit{Pure-PQC} traffic nor extrapolate our \textit{Hybrid-PQC} results to that setting.}

\subsection{Encrypted Traffic Classifiers: Representations and Learning Paradigms}

Encrypted traffic classifiers infer application, service, or activity labels from observable traffic without decrypting application payloads~\cite{SoKDecodingEnigma2025a,qingTrainingRobustClassifiers2025,deng2026enhancing}. Their effectiveness depends on both the learning architecture and the representation supplied to that architecture. Existing methods span hand-crafted flow statistics, ordered packet sequences, raw bytes, and transferable representations learned through self-supervision~\cite{shenMachineLearningPoweredEncrypted2023}. These approaches coexist because they offer different balances among information coverage, computational cost, interpretability, and data requirements. We therefore organize prior work primarily by input representation, while treating pre-training as a learning strategy layered on byte-level or structured traffic inputs.

\noindent\textbf{Aggregate statistical representations.}
Statistical methods summarize a flow using a fixed feature vector before classification. Common features include packet and byte counts, flow duration, packet-length distributions, direction ratios, burst statistics, and inter-arrival times~\cite{shenMachineLearningPoweredEncrypted2023,nguyenSurveyTechniquesInternet2008}. These features are typically combined with lightweight classifiers such as support vector machines, decision trees, random forests, or nearest-neighbor methods~\cite{hayesKFingerprintingRobust2016,van2020flowprint}. Feature selection and normalization can reduce dimensionality and improve efficiency, but they require substantial domain knowledge and careful dataset-specific tuning. CUMUL is a representative method that combines summary statistics with a sampled cumulative curve derived from direction-coded packet lengths~\cite{cumulWebsiteFingerprintingInternet2016}. The curve preserves the progression of transferred volume while maintaining a compact, fixed-dimensional representation. Statistical methods remain attractive for their efficiency and interpretability, although their performance depends strongly on manual feature design~\cite{najm2024enhanced}.

\noindent\textbf{Packet-sequence representations.}
Sequence-based methods retain packet ordering and allow a model to learn spatial or temporal dependencies directly from a trace~\cite{liuFSNetFlowSequence2019}. This approach reduces reliance on manual aggregation while preserving patterns that fixed statistical vectors may obscure. Prior work has explored recurrent sequence encoders and image-like transformations of packet sizes and arrival times~\cite{shapiraFlowPicGenericRepresentation2021,shen2021accurate}. DF uses one-dimensional convolutions to learn local motifs and hierarchical patterns from direction-coded packet traces~\cite{DeepFingerprintingUndermining2018a}. Tik-Tok extends directional modeling with timing information, enabling the classifier to capture temporal dependencies beyond packet order alone~\cite{TikTokUtilityPacket2020}. These models can represent finer traffic structure than aggregate statistics, but their inputs require explicit choices about trace length, padding, truncation, and positional alignment. Overall, sequence representations replace extensive feature construction with end-to-end learning over a defined packet-level observation surface.

\noindent\textbf{Raw-byte representations and pre-trained models.}
Raw-byte representations retain selected packet headers and encrypted payload bytes instead of reducing each packet to a small set of side-channel measurements~\cite{liu2024atvitsc,zhang2023tfe}. Early convolutional methods showed that discriminative traffic features can be learned directly from byte sequences without manually defined flow statistics~\cite{wangMalwareTrafficClassification2017}. This richer observation surface also requires careful preprocessing because identifiers, protocol fields, packet selection, and truncation can introduce dataset-specific artifacts~\cite{SoKDecodingEnigma2025a}. Pre-training changes how these representations are learned rather than defining a separate input category. It first learns reusable patterns from large unlabeled traffic corpora and then adapts the resulting encoder to labeled downstream tasks~\cite{peng2025bottom,wang2024netmamba,zhouTrafficFormerEfficient2025}. ET-BERT pre-trains a Transformer over contextualized datagram representations, while YaTC uses masked autoencoding to learn a multi-level flow representation~\cite{ETBERTContextualizedDatagram2022a,YatcAnotherTrafficClassifier2023a}. These models combine information-rich inputs with self-supervised learning, but the quality of their transfer still depends on the relationship between pre-training and downstream distributions~\cite{zhaoSweetDangerSugar2025}. LLM-based approaches further extend this learning strategy through traffic-specific tokenization, instruction tuning, and open-set learning~\cite{cui2025trafficllm,ginigeTrafficGPT2024,zhao2026robust}. Overall, raw-byte processing determines what information is exposed, whereas pre-training determines how reusable features are acquired from that information.

\subsection{Distribution Drift in Encrypted Traffic Classification}

\begin{table*}[t]
\centering
\caption{Taxonomy of Distribution-Drift Sources in Encrypted Traffic Classification}
\label{tab:drift_taxonomy}
\small
\renewcommand{\arraystretch}{1.2}
\begin{tabularx}{\textwidth}{@{} >{\RaggedRight}p{2.2cm} >{\RaggedRight}X >{\RaggedRight}X >{\RaggedRight}X @{}}
\toprule
\textbf{Drift Type} & 
\textbf{Sources of Variation} & 
\textbf{Prior Focus} & 
\textbf{Control or Scope in This Study} \\ 
\midrule
Temporal Drift & 
Web content, server or CDN updates, third-party resources, and collection time. & 
Cross-time degradation, resampling, and model updating~\cite{fanResAwareCrossEnvironment2026,chen2025drift}. & 
Adjacent, alternating Non-PQC and PQC collections limit temporal confounding. \\ 
\addlinespace[0.6em]

Network Drift & 
Vantage points, ISPs or ASNs, routing, RTT, loss, and bandwidth. & 
Cross-vantage and open-world generalization~\cite{fanResAwareCrossEnvironment2026,shadbehRealityCheckTor2026}. & 
A fixed vantage point controls path and access-network variation. \\ 
\addlinespace[0.6em]

Client Drift & 
Browsers, operating systems, TLS libraries, transport stacks, caching, and loading strategies. & 
Robustness across clients and software stacks~\cite{fanResAwareCrossEnvironment2026}. & 
The client platform, protocol stack, cache state, and scripts remain fixed. \\ 
\addlinespace[0.6em]

Application Drift & 
Version updates, API migrations, and server-side logic. & 
Self-evolving classification, unknown-pattern discovery, and open-world adaptation~\cite{yuanM3SUPD2025,xianUDFS2025}. & 
Website labels and access tasks remain unchanged across collections. \\ 
\addlinespace[0.6em]

Pipeline Drift & 
Collection, segmentation, labels, sanitization, truncation, and data splits. & 
Benchmark comparability, reproducibility, and dataset stability~\cite{SoKDecodingEnigma2025a,akbariOneTaskRule2025,soukupDriftBasedDataset2025}. & 
Labels, splits, and metrics are unified; native model inputs are disclosed. \\ 
\addlinespace[0.6em]

Protocol Drift & 
TLS or QUIC versions, extensions, key exchange, and transport migration. & 
Protocol measurement, handshake overhead, and implementation fingerprinting~\cite{zhang2023application,gomezCambroneroLayeredTLS2026,mallickClassifyingPQCImplementations2025}. & 
PQC key establishment is varied while other collection factors are controlled. \\ 

\bottomrule
\end{tabularx}
\end{table*}

\textit{Distribution drift} refers to a mismatch between the traffic distributions observed during model training and deployment~\cite{chen2025drift,soukupDriftBasedDataset2025,deng2026enhancing}. Such mismatches can arise from collection time, network paths, client platforms, service evolution, data pipelines, or the protocol stack itself. They can therefore undermine deployment reliability even when a classifier performs well under mismatched experimental conditions. Table~\ref{tab:drift_taxonomy} summarizes these sources and distinguishes how each is controlled or studied in this work.

Prior studies show that encrypted traffic classifiers often learn regularities tied to their collection environments. Cross-environment evaluations report degradation across time, vantage points, and client settings~\cite{fanResAwareCrossEnvironment2026,shadbehRealityCheckTor2026}. Other work examines application evolution, unknown-pattern discovery, and dataset instability~\cite{chen2025drift,xianUDFS2025,soukupDriftBasedDataset2025}. Wickramasinghe \emph{et al.} further show that inputs, sanitization and sample granularity can materially affect reported conclusions~\cite{SoKDecodingEnigma2025a}. Comparatively less work isolates the evolution of the protocol stack as an independent source of distribution drift.

Within this taxonomy, we define \textit{protocol drift} as shifts in traffic distribution resulting from the evolution or deployment of protocol mechanisms. Unlike temporal, network, or application drift, protocol drift alters the traffic-generation process even when labels, clients, vantage points, and collection pipelines are held constant. It can arise from changes to TLS extensions, key-establishment algorithms, encrypted handshake mechanisms~\cite{rfc9849}, or migration to QUIC and HTTP/3~\cite{cheng2026star}. Prior measurement studies show that protocol design and implementation choices can alter observable traffic patterns and performance characteristics~\cite{zhang2023application,gomezCambroneroLayeredTLS2026,mallickClassifyingPQCImplementations2025}. This paper examines PQC evolution as a concrete instance of protocol drift, focusing on the TLS~1.3 transition from X25519 to hybrid \texttt{\detokenize{X25519MLKEM768}} key establishment. The resulting changes to the handshake material may affect packet segmentation, timing, and the byte-level representations used as inputs to encrypted traffic classifiers.

%% file: Tex/3-Analysis.tex
\section{PQC-Induced Protocol Drift in TLS 1.3 Traffic}
\label{sec:analy}

\subsection{Comparison of TLS 1.3 Key-Establishment Modes}

\subsubsection{Traditional (Non-PQC) Key Establishment}

\begin{figure}[ht]
    \centering
    \includegraphics[width=\columnwidth]{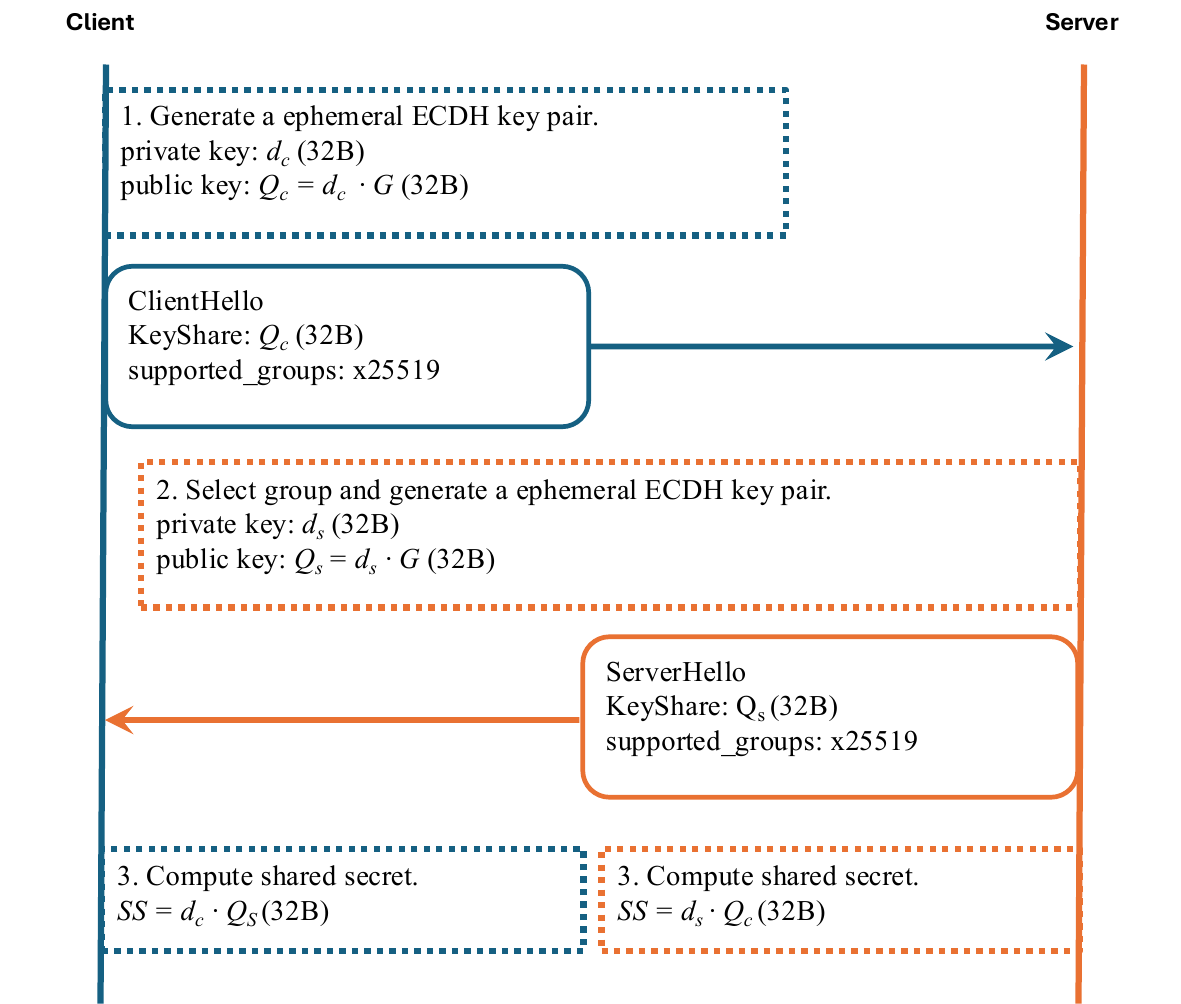}
    \caption{Traditional (Non-PQC) X25519 key establishment in TLS~1.3.}
    \label{pic_tls13_traditional}
\end{figure}

The Traditional (Non-PQC) mode considered in this paper uses ephemeral X25519 Diffie--Hellman for key establishment in TLS~1.3~\cite{rfc9846TLS13}. During the handshake, the client and server exchange ephemeral key shares and independently derive the same shared secret. The TLS~1.3 key schedule incorporates this secret through HKDF-based extraction and expansion to derive handshake and application traffic secrets and, ultimately, traffic keys~\cite{rfc9846TLS13,RFC5869HMACbasedExtractandExpandKey2010}.

In the \texttt{ClientHello}, the client advertises acceptable named groups in the \texttt{supported\_groups} extension and typically supplies a \texttt{KeyShareEntry} for one or more of these groups in the \texttt{key\_share} extension. The server selects a mutually supported group and returns the corresponding key share in the \texttt{ServerHello}. If the client has not supplied a key share for a group selected by the server, the server can instead send a \texttt{HelloRetryRequest}. TLS~1.3 supports elliptic-curve groups, such as \texttt{secp256r1} and \texttt{x25519}, as well as finite-field Diffie--Hellman groups. Our Traditional baseline uses \texttt{x25519}, whose key share is 32 bytes~\cite{rfc9846TLS13}.

As illustrated in Fig.~\ref{pic_tls13_traditional}, we express X25519 using conventional ECDH notation and let $G$ denote the fixed base point. The client generates an ephemeral private scalar $d_c$, computes the corresponding key share $Q_c=d_c\cdot G$, and sends $Q_c$ in the \texttt{ClientHello}. After selecting \texttt{x25519}, the server generates $d_s$, computes $Q_s=d_s\cdot G$, and returns $Q_s$ in the \texttt{ServerHello}. The client derives $SS_{\mathrm{X25519}}=d_c\cdot Q_s$, while the server derives $SS_{\mathrm{X25519}}=d_s\cdot Q_c$. Both computations yield $SS_{\mathrm{X25519}}=d_c\cdot d_s\cdot G$, which is then supplied to the TLS~1.3 key schedule.

\subsubsection{Hybrid-PQC Key Establishment}

\begin{figure}[ht]
    \centering
    \includegraphics[width=\columnwidth]{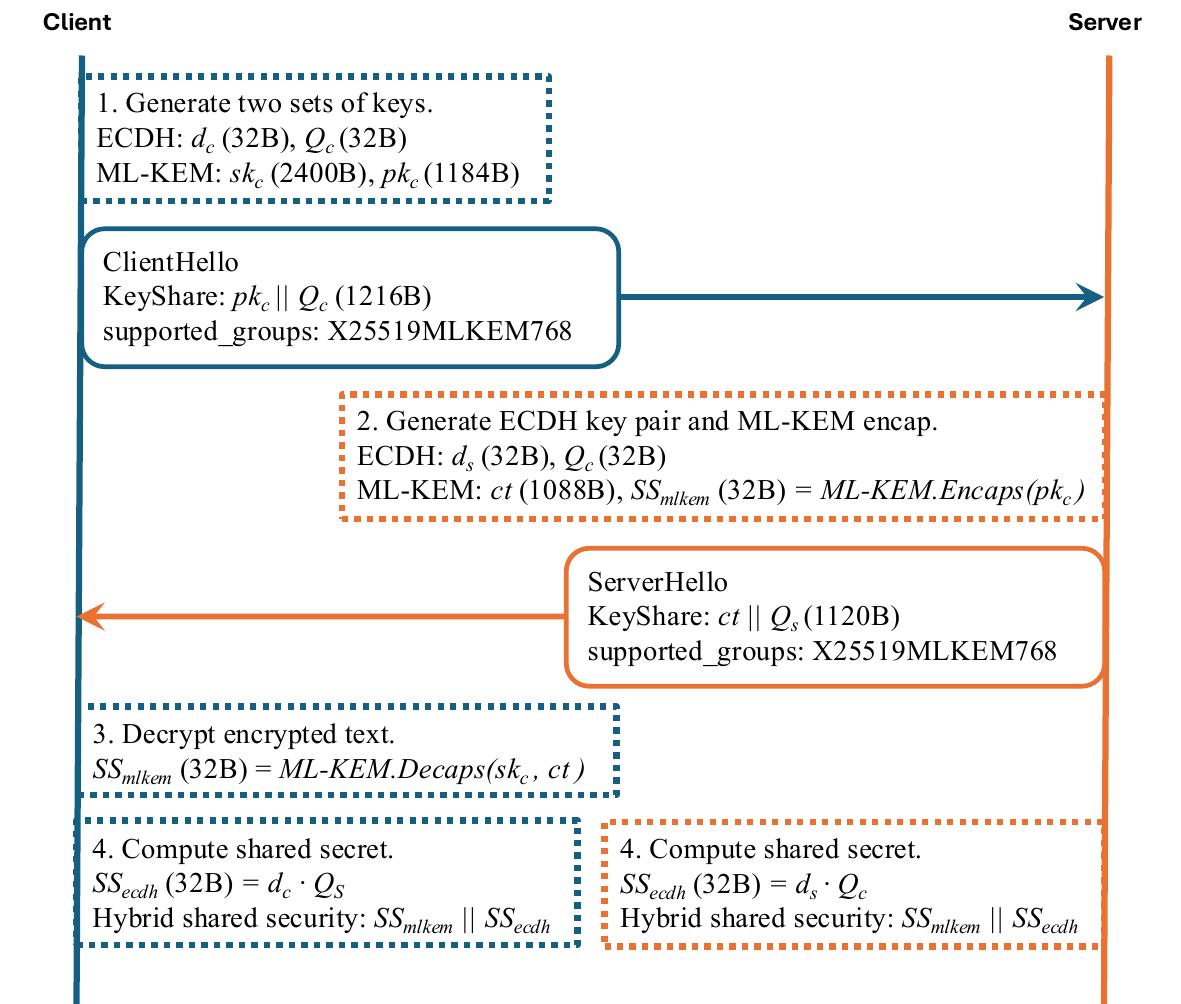}
    \caption{Hybrid-PQC \texttt{X25519MLKEM768} key establishment in TLS~1.3.}
    \label{pic_tls13_hybrid}
\end{figure}

RFC~9954 defines a framework for hybrid key exchange in TLS~1.3, in which a traditional key-exchange component and a post-quantum KEM component are represented by a single named group~\cite{RFC9954HybridKeyExchange2026}. The component \texttt{key\_exchange} values are concatenated within each \texttt{KeyShareEntry}, and the concatenated shared secret is supplied to the TLS~1.3 key schedule in place of the conventional (EC)DHE secret. This construction is designed to preserve security as long as at least one component remains secure.

The \texttt{draft-ietf-tls-ecdhe-mlkem-05} defines three hybrid named groups that combine an elliptic-curve Diffie--Hellman group with an ML-KEM parameter set~\cite{kwiatkowskiPostquantumHybridECDHEMLKEM2026}. Among them, \texttt{X25519MLKEM768} is marked \texttt{Recommended=Y} and is the Hybrid-PQC configuration studied in this paper. Its client \texttt{KeyShareEntry} contains a 1184-byte ML-KEM-768 encapsulation key followed by a 32-byte X25519 key share, for a total of 1216 bytes. Its server entry contains a 1088-byte ML-KEM-768 ciphertext followed by a 32-byte X25519 key share, for a total of 1120 bytes.

As illustrated in Fig.~\ref{pic_tls13_hybrid}, the client generates an X25519 key pair $(d_c,Q_c)$ and an ML-KEM key pair $(pk_c,sk_c)$, and sends $pk_c\parallel Q_c$ in the \texttt{ClientHello}. The server generates $(d_s,Q_s)$, runs \textsf{Encaps} with $pk_c$ to obtain $(ct,SS_{\mathrm{MLKEM}})$, and returns $ct\parallel Q_s$ in the \texttt{ServerHello}. The client runs \textsf{Decaps} with $sk_c$ and $ct$ to recover $SS_{\mathrm{MLKEM}}$. Both endpoints also derive $SS_{\mathrm{X25519}}$ from their X25519 key pairs. The two component secrets are concatenated in the specified order:
\begin{equation}
SS_{\mathrm{hybrid}}=SS_{\mathrm{MLKEM}}\parallel SS_{\mathrm{X25519}},
\end{equation}
and the resulting 64-byte hybrid shared secret is supplied to the TLS~1.3 key schedule.

\subsubsection{Pure-PQC Key Establishment}

\begin{figure}[ht]
    \centering
    \includegraphics[width=\columnwidth]{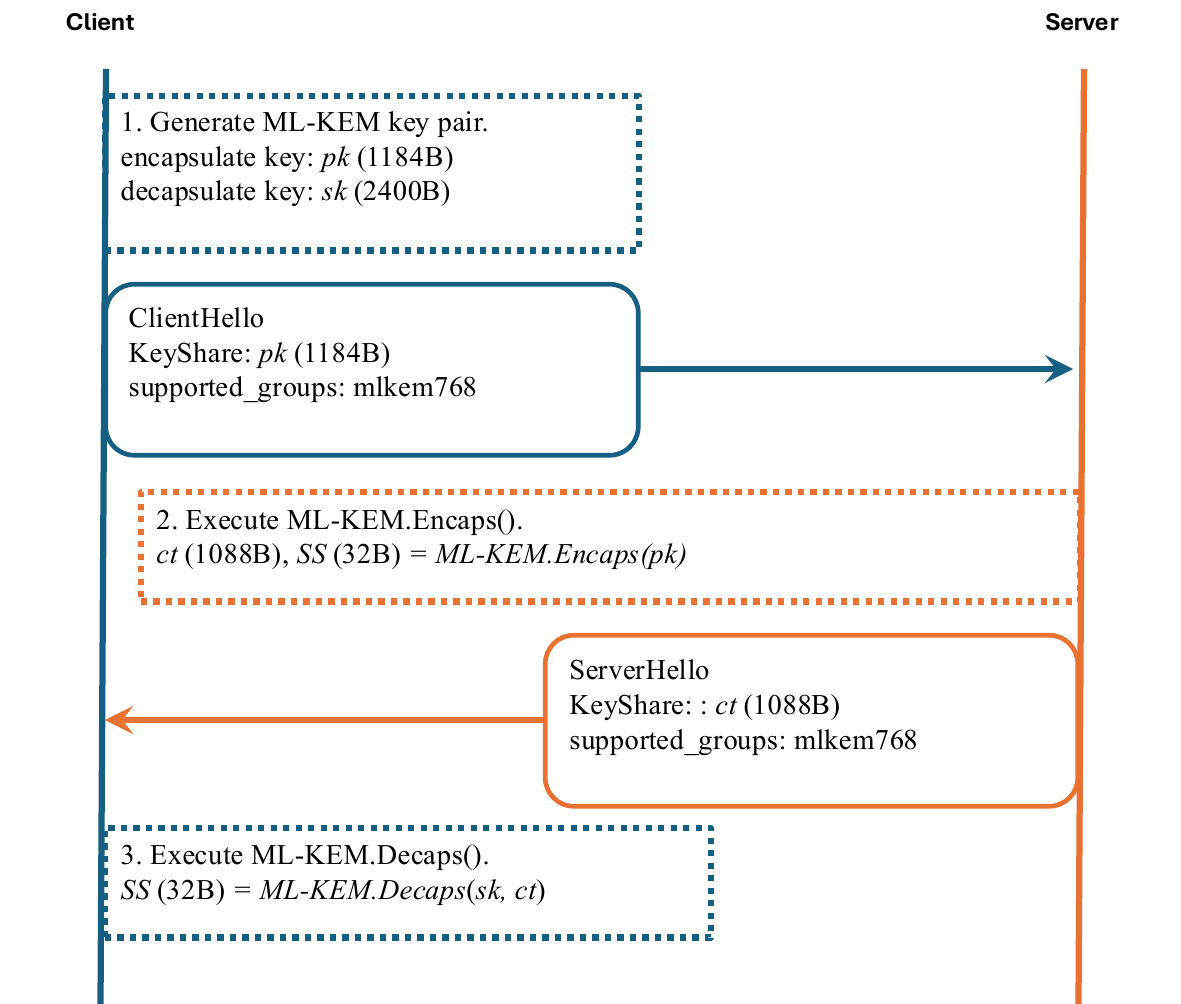}
    \caption{Pure-PQC \texttt{MLKEM768} key establishment in TLS~1.3.}
    \label{pic_tls13_pq}
\end{figure}

Pure-PQC key establishment removes the conventional (EC)DHE exchange and uses ML-KEM within the standard TLS~1.3 handshake flow. This mechanism and its associated named groups are specified in the IETF draft \texttt{draft-ietf-tls-mlkem-09}~\cite{pureMLKEMPostQuantumKey2026a}. These groups are \texttt{MLKEM512}, \texttt{MLKEM768}, and \texttt{MLKEM1024}, corresponding to the standardized ML-KEM parameter sets~\cite{nistFIPS203MLKEM2024}. We use \texttt{MLKEM768} for the protocol-level comparison because it matches the KEM parameter set in \texttt{X25519MLKEM768}; consistent with the scope defined in Section~\ref{sec:back}, Pure-PQC is not included in the empirical evaluation.

As illustrated in Fig.~\ref{pic_tls13_pq}, the client generates an ML-KEM encapsulation/decapsulation key pair $(pk,sk)$ and sends the encapsulation key $pk$ in the \texttt{key\_share} extension of the \texttt{ClientHello}. The server runs \textsf{Encaps} with $pk$ to produce a ciphertext $ct$ and a shared secret $SS_{\mathrm{MLKEM}}$, then returns $ct$ in the \texttt{ServerHello}. The client runs \textsf{Decaps} with $sk$ and $ct$ to recover the same shared secret. This secret is supplied directly to the TLS~1.3 key schedule in place of an (EC)DHE shared secret.

\subsubsection{Comparison of Key-Establishment Modes}

\begin{table*}[htbp]
\centering
\caption{Comparison of TLS~1.3 key-establishment modes.}
\label{tab:paradigm_multidim_comparison}
\small
\renewcommand{\arraystretch}{1.2}
\begin{tabularx}{\textwidth}{@{} >{\raggedright\arraybackslash}m{2.2cm} >{\raggedright\arraybackslash}m{2.2cm} >{\raggedright\arraybackslash}m{2.6cm} >{\raggedright\arraybackslash}X >{\raggedright\arraybackslash}X @{}}
\toprule
\textbf{Mode} &
\textbf{Composition} &
\textbf{Client / Server KeyShare Size} &
\textbf{Cryptographic Operations} &
\textbf{Characteristics} \\
\midrule

Traditional (Non-PQC) &
\texttt{X25519} &
32 B / 32 B &
X25519 key generation + shared-secret computation &
Compact key shares with no post-quantum component \\

\addlinespace[0.6em]

Hybrid-PQC &
\makecell[l]{\texttt{X25519 + } \\ \texttt{ML-KEM-768}} &
1216 B / 1120 B &
X25519 + ML-KEM (\textsf{KeyGen}, \textsf{Encaps}, \textsf{Decaps}) &
Retains X25519 and adds ML-KEM, increasing the KeyShare sizes \\

\addlinespace[0.6em]

Pure-PQC &
\texttt{ML-KEM-768} &
1184 B / 1088 B &
ML-KEM (\textsf{KeyGen}, \textsf{Encaps}, \textsf{Decaps}) &
Uses standalone ML-KEM without a traditional component \\

\bottomrule
\end{tabularx}
\end{table*}

Table~\ref{tab:paradigm_multidim_comparison} compares the three TLS~1.3 key-establishment modes by their \texttt{KeyShareEntry} sizes and cryptographic operations. The Traditional (Non-PQC) mode exchanges 32-byte X25519 key shares in both directions. Hybrid-PQC (\texttt{X25519MLKEM768}) increases the client and server entries to 1216 and 1120 bytes, respectively, by combining X25519 with ML-KEM-768. Pure-PQC (\texttt{MLKEM768}) removes the X25519 component and uses an 1184-byte encapsulation key from the client and a 1088-byte ciphertext from the server. Thus, Pure-PQC reduces each entry by 32 bytes relative to Hybrid-PQC, but both PQC modes carry substantially larger key-establishment payloads than the Traditional baseline.

These larger key-establishment payloads increase the sizes of the \texttt{ClientHello} and \texttt{ServerHello}. When a serialized handshake flight exceeds the negotiated TCP Maximum Segment Size (MSS), the TCP stack distributes its bytes across multiple segments. The additional segments can increase the observable packet count and incur additional transmission and retransmission costs.

Together, these protocol-level changes modify the observable side-channel surface of encrypted traffic. Larger \texttt{KeyShareEntry} values can change packet-length sequences, burst structure, timing, TLS-record boundaries, and byte-level alignment---all of which are inputs to the traffic classifiers studied in this paper. The next subsection analyzes these effects across four feature dimensions.

\subsection{PQC-Induced Changes Across Four Feature Dimensions}

We represent the observable traffic trace generated by a single HTTPS website visit as a 4-tuple:
\begin{equation}
X = (L, D, T, P)
\end{equation}
Here, $L$ is the packet-length sequence, $D$ is the corresponding packet-direction sequence, and $T$ contains packet timestamps or inter-arrival times (IATs). The fourth component, $P$, denotes a generalized byte-level packet representation. In this paper, $P$ may include sanitized network and transport headers, TLS handshake bytes, and encrypted application records, but it excludes plaintext application-layer content. This formulation covers the native inputs of the classifier families evaluated later. \textsf{CUMUL}, \textsf{DF}, and \textsf{Tik-Tok} derive their inputs primarily from combinations of $L$, $D$, and $T$~\cite{cumulWebsiteFingerprintingInternet2016,DeepFingerprintingUndermining2018a,TikTokUtilityPacket2020}. By comparison, \textsf{ET-BERT} and \textsf{YaTC} consume richer byte-oriented representations captured by $P$~\cite{ETBERTContextualizedDatagram2022a,YatcAnotherTrafficClassifier2023a}. The 4-tuple therefore separates compact side-channel sequences from selected packet bytes without assuming that all classifiers observe the same input fields.

Relative to Traditional (Non-PQC) X25519 key establishment, Hybrid-PQC adds an ML-KEM-768 encapsulation key to the \texttt{ClientHello} and an ML-KEM-768 ciphertext to the \texttt{ServerHello}. As summarized in Table~\ref{tab:paradigm_multidim_comparison}, these additions originate in the \texttt{key\_share} extension but propagate through TLS record construction and TCP segmentation. Their effects differ across feature dimensions and depend partly on each classifier's preprocessing pipeline.

\noindent\textbf{Packet length ($L$).}
The ML-KEM-768 encapsulation key and ciphertext are substantially larger than the 32-byte X25519 key shares used in the Traditional mode, directly increasing the early handshake payloads. This increase changes the lengths of packets carrying the \texttt{ClientHello} and \texttt{ServerHello}, as well as the total byte volume of the handshake. When the expanded messages cross TLS record or TCP segment boundaries, they can also shift the positions of subsequent packets within a fixed-length trace. These changes affect both packet-length distributions and cumulative byte curves, including the representation used by \textsf{CUMUL}.

\noindent\textbf{Packet direction ($D$).}
Hybrid-PQC does not change the logical directions of the \texttt{ClientHello} and \texttt{ServerHello}. Instead, larger handshake messages may be carried in more TCP segments. If preprocessing retains these segments and their acknowledgments, the number and positions of client-to-server and server-to-client packets can change within the first $N$ observations. Even when pure ACKs are removed, longer runs of same-direction data packets can alter burst boundaries and uplink--downlink patterns. The effect on $D$ therefore depends on both TCP segmentation and the packet-filtering policy used by the classifier.

\noindent\textbf{Timing ($T$).}
Timing changes can arise from both transmission and cryptographic processing. Additional handshake bytes require more serialization and may introduce extra segments, acknowledgments, or retransmissions under constrained network conditions. Hybrid-PQC also adds ML-KEM \textsf{KeyGen}, \textsf{Encaps}, and \textsf{Decaps} operations to the X25519 computation. Their relative cost depends on the implementation, hardware, and available acceleration, so it should not be assumed to exceed X25519 in every deployment. Together, these factors may shift packet IATs and handshake-phase duration, particularly on bandwidth-limited paths or resource-constrained endpoints. Unlike the deterministic increase in keying-material size, the magnitude of timing drift is therefore environment-dependent.

\noindent\textbf{Byte-level packet representation ($P$).}
Hybrid-PQC changes both the content and length of the \texttt{key\_share} extension. The client entry concatenates the ML-KEM-768 encapsulation key with the client's 32-byte X25519 ephemeral key share. The server entry concatenates the ML-KEM-768 ciphertext with the server's 32-byte X25519 ephemeral key share. These fields change the byte sequence and offsets within the corresponding handshake messages. Larger messages may also cross TLS record and TCP segment boundaries at different positions, shifting the alignment of later bytes. Under fixed or truncated observation windows, a classifier may therefore capture different portions of the handshake and subsequent encrypted records. Pre-trained encoders such as \textsf{ET-BERT} and \textsf{YaTC} can consequently receive familiar protocol fields at new offsets or omit content that appeared within the Non-PQC window. This constitutes a representation-level distribution shift rather than a change to the semantics of the encrypted application content.

Taken together, Hybrid-PQC changes $L$ and $P$ directly through the expanded \texttt{key\_share} values, while its effects on $D$ and $T$ arise through segmentation, scheduling, and implementation-dependent processing. This distinction links the protocol mechanisms described above to the testable hypotheses and feature-control experiments developed in the following sections.

\subsection{Hypotheses on Classifier Robustness under PQC-Induced Protocol Drift}

Building on the 4-tuple representation above, we formulate three testable hypotheses about the cross-domain robustness of encrypted traffic classifiers under PQC-induced protocol drift. Together, these hypotheses provide the analytical basis for the research questions examined later, linking protocol-level changes to classifier robustness during PQC migration.

\par\smallskip
\noindent\textbf{H1: Packet-Length Drift.}\enspace Hybrid-PQC expands the KEM payloads carried during the TLS handshake, changing early packet lengths, positional offsets, and cumulative traffic profiles. The packet-length sequence $L$ and its derived statistics (e.g., cumulative curves and statistical moments) are primary inputs to encrypted traffic classifiers. We therefore hypothesize that models relying heavily on $L$ will exhibit substantial performance degradation when trained on Non-PQC traffic and evaluated on PQC traffic.

\par\smallskip
\noindent\textbf{H2: Directional and Temporal Drift.}\enspace Hybrid-PQC may also alter the direction sequence $D$ and timing sequence $T$. Larger handshake messages can require additional TCP segments and, depending on preprocessing, change the positions of packets and acknowledgments within the first $N$ observations, thereby modifying uplink--downlink patterns. 
Hybrid-PQC introduces computational overhead from ML-KEM \textsf{KeyGen}, \textsf{Encaps}, and \textsf{Decaps}, as well as transmission overhead from serializing larger handshake messages. Together, these factors may alter packet inter-arrival time (IAT) distributions and handshake completion time. We therefore hypothesize that classifiers using temporal or combined direction--timing features will degrade across the Non-PQC and PQC domains, with larger effects on high-latency paths or resource-constrained endpoints.

\par\smallskip
\noindent\textbf{H3: Robustness of Pre-Trained Byte-Level Representations.}\enspace Pre-trained traffic models such as \textsf{ET-BERT} and \textsf{YaTC} may partially reduce PQC-induced cross-domain degradation by capturing reusable byte-level structures in $P$. They may therefore be more robust than classifiers based primarily on $L$, $D$, and $T$, although fixed input windows and preprocessing choices constrain this advantage. \textsf{ET-BERT}'s input horizon may omit contextual bytes as handshake messages expand, whereas \textsf{YaTC}'s extraction window may capture shifted byte ranges when TLS-record boundaries change. If their pre-training corpora do not include Hybrid-PQC traffic, whether they can preserve their Non-PQC baseline performance remains uncertain.

%% file: Tex/4-Dataset.tex
\section{Building a PQC-Aware Encrypted Traffic Benchmark}
\label{sec:dataset}

Section~\ref{sec:analy} showed how Hybrid-PQC key establishment can alter the observable traffic features used by encrypted traffic classifiers. To evaluate these effects empirically, we build a PQC-aware encrypted traffic benchmark in three stages. We first measure TLS key-establishment deployment across 2,060 Tranco-ranked websites. We then collect paired Non-PQC and Hybrid-PQC traffic for 195 common websites. Finally, we construct deployment-ratio test sets that vary the Hybrid-PQC proportion while holding the label space and visit-level partitions fixed.

\subsection{Measuring TLS Key-Establishment Deployment}

To characterize TLS key-establishment deployment and identify candidate websites for paired traffic collection, we analyze 2,060 websites drawn from the Tranco ranking~\cite{pochatTrancoResearchOrientedTop}. For each website, we record the negotiated TLS version and use Algorithm~\ref{alg:keyshare_target_sni} to extract the distribution of \texttt{key\_share} groups selected in \texttt{ServerHello}. The target SNI set includes both the base domain and its \texttt{www}-prefixed variant.

Algorithm~\ref{alg:keyshare_target_sni} processes each packet capture at the granularity of individual TLS sessions. It first identifies sessions containing both \textit{ClientHello} and \textit{ServerHello}. The client SNI is then compared with the target set, and each matching session contributes its \texttt{ServerHello} \texttt{key\_share} group to the corresponding count. Separate counters record matched sessions without a \texttt{key\_share} group, sessions with non-target SNIs, and incomplete handshakes. These counters retain the filtering outcomes needed to interpret the final group distribution. The resulting statistics characterize the key-establishment mechanisms observed for each website and guide the selection of websites used to construct the benchmark.

\begin{algorithm}[ht]
    \caption{Extraction of Negotiated TLS Key-Share Distributions}
    \label{alg:keyshare_target_sni}
    \footnotesize 
    \SetAlgoNoLine
    \LinesNumbered

    \SetKwInOut{Input}{Input}
    \SetKwInOut{Output}{Output}
    \SetKwInOut{Notation}{Notations}

    \SetKwFor{For}{for}{do}{end\ for}
    \SetKwIF{If}{ElseIf}{Else}{if}{then}{else if}{else}{end\ if}

    \Input{Packet-capture dataset \texttt{dataset}; target SNI set \texttt{target\_snis}}
    \Output{Dictionary \texttt{keyshare\_stats}}

    \Notation{
        \texttt{ks\_dist} : \texttt{keyshare\_stats["key\_share\_groups"]}\\
        \texttt{matched} \hspace{0.5mm} : \texttt{keyshare\_stats["matched\_count"]}\\
        \texttt{no\_ks} \hspace{0.5mm} : \texttt{keyshare\_stats["no\_key\_share"]}\\
        \texttt{unmatched} : \texttt{keyshare\_stats["sni\_unmatched"]}\\
        \texttt{missing} \hspace{0.5mm} : \texttt{keyshare\_stats["missing\_handshake"]}
    }
    \smallskip

    \setcounter{AlgoLine}{0}

    Initialize \texttt{keyshare\_stats}\;
    \For{\textup{each} capture\_file \textbf{in} \texttt{dataset}}{
        $sessions \leftarrow$ Extract TLS sessions from \textit{capture\_file}\;
        \For{\textup{each} session \textbf{in} sessions}{
            $has\_ch \leftarrow$ Check whether \textit{ClientHello} exists\;
            $has\_sh \leftarrow$ Check whether \textit{ServerHello} exists\;
            \uIf{$has\_ch$ \textbf{and} $has\_sh$}{
                $sni \leftarrow$ Extract \texttt{server\_name} from \textit{ClientHello}\;
                \uIf{$sni \in \texttt{target\_snis}$}{
                    \texttt{matched} $\leftarrow$ \texttt{matched} + 1\;
                    $ks \leftarrow$ Extract \texttt{key\_share} group from \textit{ServerHello}\;
                    \uIf{$ks$ \textbf{is not} NULL}{
                        \texttt{ks\_dist}[$ks$] += 1\;
                    }
                    \Else{
                        \texttt{no\_ks} $\leftarrow$ \texttt{no\_ks} + 1\;
                    }
                }
                \Else{
                    \texttt{unmatched} $\leftarrow$ \texttt{unmatched} + 1\;
                }
            }
            \Else{
                \texttt{missing} $\leftarrow$ \texttt{missing} + 1\;
            }
        }
    }
    \Return \texttt{keyshare\_stats}\;
\end{algorithm}

\begin{figure}[htbp]
\centering
\begin{tikzpicture}[every node/.append style={font=\small}]
\definecolor{cNoReply}{HTML}{999999}
\definecolor{cTLS12}{HTML}{009E73}
\definecolor{cHybrid}{HTML}{0072B2}
\definecolor{cTrad}{HTML}{D55E00}
\def\R{2.2}
\fill[fill=cNoReply!25, pattern=dots, pattern color=cNoReply!90!black,
      draw=white, line width=1.2pt]
  (0,0) -- ({90}:{\R}) arc ({90}:{85.32}:{\R}) -- cycle;
\fill[fill=cTLS12!25, pattern=north east lines, pattern color=cTLS12!80!black,
      draw=white, line width=1.2pt]
  (0,0) -- ({85.32}:{\R}) arc ({85.32}:{50.04}:{\R}) -- cycle;
\fill[fill=cHybrid!25, pattern=horizontal lines, pattern color=cHybrid!80!black,
      draw=white, line width=1.2pt]
  (0,0) -- ({50.04}:{\R}) arc ({50.04}:{-131.04}:{\R}) -- cycle;
\fill[fill=cTrad!25, pattern=crosshatch, pattern color=cTrad!80!black,
      draw=white, line width=1.2pt]
  (0,0) -- ({-131.04}:{\R}) arc ({-131.04}:{-270}:{\R}) -- cycle;
\node[fill=white, inner sep=1pt, rounded corners=1pt, font=\small\bfseries]
  at ({67.68}:{0.60*\R}) {9.8\%};
\node[fill=white, inner sep=1pt, rounded corners=1pt, font=\small\bfseries]
  at ({-40.5}:{0.60*\R}) {50.3\%};
\node[fill=white, inner sep=1pt, rounded corners=1pt, font=\small\bfseries]
  at ({159.48}:{0.60*\R}) {38.6\%};
\filldraw[black] ({87.66}:{\R}) circle (0.8pt);
\draw[thick, black] ({87.66}:{\R}) -- ({87.66}:{2.25}) -- (1.7,2.25)
  node[right, black] {1.3\%};
\begin{scope}[yshift=-0.2cm]
\def\s{0.38}

\fill[fill=cNoReply!60, pattern=dots, pattern color=cNoReply!90!black, draw=black!40]
  (-3.0,-2.6) rectangle ({-3.0+\s},{-2.6+\s});
\node[right] at ({-3.0+\s+0.1},{-2.6+\s/2}) {No Reply};

\fill[fill=cHybrid!60, pattern=horizontal lines, pattern color=cHybrid!80!black, draw=black!40]
  (-3.0,-3.3) rectangle ({-3.0+\s},{-3.3+\s});
\node[right] at ({-3.0+\s+0.1},{-3.3+\s/2}) {TLS 1.3 Hybrid-PQC};

\fill[fill=cTLS12!60, pattern=north east lines, pattern color=cTLS12!80!black, draw=black!40]
  (1.0,-2.6) rectangle ({1.0+\s},{-2.6+\s});
\node[right] at ({1.0+\s+0.1},{-2.6+\s/2}) {TLS 1.2};

\fill[fill=cTrad!60, pattern=crosshatch, pattern color=cTrad!80!black, draw=black!40]
  (1.0,-3.3) rectangle ({1.0+\s},{-3.3+\s});
\node[right] at ({1.0+\s+0.1},{-3.3+\s/2}) {TLS 1.3 Non-PQC};
\end{scope}
\end{tikzpicture}
\caption{Distribution of negotiated TLS versions and TLS~1.3 key-establishment modes across 2,060 Tranco-ranked websites. TLS~1.3 Non-PQC and Hybrid-PQC are identified by server-side \texttt{key\_share} values of 29 (\textsf{X25519}) and 4588 (\textsf{X25519MLKEM768}), respectively.}
\label{fig:tls_distribution}
\end{figure}
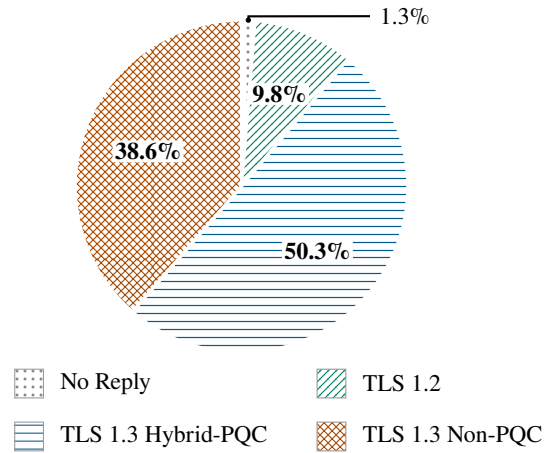

For TLS~1.3 connections, the group selected in \texttt{ServerHello} identifies the key-establishment mechanism used for that connection. Fig.~\ref{fig:tls_distribution} reports the observed TLS version and key-establishment distribution. Among the 2,060 websites, 50.3\% negotiated Hybrid-PQC TLS~1.3 key establishment, while 38.6\% negotiated Traditional (Non-PQC) TLS~1.3 key establishment. A further 9.8\% used TLS~1.2, and 1.3\% returned no reply. Hybrid-PQC was therefore the largest category in the measured sample and provided a sufficiently large candidate pool for the paired traffic collection described next.

\subsection{Constructing Paired Non-PQC and Hybrid-PQC Traffic Datasets}

The deployment results obtained with Algorithm~\ref{alg:keyshare_target_sni} and summarized in Fig.~\ref{fig:tls_distribution} show that the observed Hybrid-PQC TLS~1.3 connections selected the \texttt{\detokenize{X25519MLKEM768}} group. We therefore selected the Top 200 websites that support this group as candidates for paired traffic collection and used them to construct the \textsf{Non-PQC Website Dataset} and the \textsf{PQC Website Dataset}. Throughout this paper, the \textsf{PQC Website Dataset} refers specifically to traffic collected with Hybrid-PQC \texttt{\detokenize{X25519MLKEM768}} key establishment, rather than \textit{Pure-PQC} key establishment.

For each candidate website, we used Mozilla Firefox to automate full-page visits under two cryptographic configurations. We toggled the \texttt{security.tls.enable\_kyber} preference to enable or disable Hybrid-PQC key establishment, thereby collecting Hybrid-PQC and Traditional (Non-PQC) traffic, respectively. We performed 120 complete page loads per website under each configuration to maintain balanced per-website sample counts across the two cryptographic domains. Each resulting packet trace corresponds to a single independent, isolated website visit.

We reapplied Algorithm~\ref{alg:keyshare_target_sni} to the collected packet captures and verified that all retained samples in the \textsf{PQC Website Dataset} negotiated \texttt{\detokenize{X25519MLKEM768}} (code point 4588), whereas all retained samples in the \textsf{Non-PQC Website Dataset} negotiated \texttt{X25519} (code point 29). Dataset membership is therefore determined by the group selected in \texttt{ServerHello}, rather than by the intended browser configuration alone.

From each validated packet capture, we extracted the TCP streams associated with the target SNIs to form the two benchmark datasets. Because a modern page load may span multiple TCP connections, we partition the data by complete website visit rather than by individual TCP stream. All streams extracted from the same visit are assigned to the same split, preventing related traffic from crossing the training, validation, and test partitions.

Network and collection failures prevented four candidate websites from yielding valid captures under both configurations. We retained the remaining 195 websites as the shared label set for the two datasets and for the closed-world evaluation in Section~\ref{sec:evalu}.

\subsection{Constructing Deployment-Ratio Test Sets}

To model gradual Hybrid-PQC deployment, we construct four Mixed test sets from the paired test partitions of the \textsf{Non-PQC Website Dataset} and the \textsf{PQC Website Dataset}. Together with the Non-PQC and Hybrid-PQC endpoint test sets, they form six deployment-ratio settings. Let $\alpha$ denote the proportion of Hybrid-PQC samples in a test set:

\begin{equation}
\alpha =
\frac{N_{\mathrm{Hybrid\text{-}PQC}}}
{N_{\mathrm{Non\text{-}PQC}} + N_{\mathrm{Hybrid\text{-}PQC}}},
\end{equation}

where $N_{\mathrm{Hybrid\text{-}PQC}}$ and $N_{\mathrm{Non\text{-}PQC}}$ denote the numbers of samples drawn from the corresponding cryptographic domains. We evaluate the following six settings:

\begin{equation}
\alpha \in \{0\%, 20\%, 40\%, 60\%, 80\%, 100\%\}.
\end{equation}

\begin{table*}[htbp]
\centering
\caption{Overview of the encrypted traffic classifiers evaluated under PQC-induced protocol drift.}
\label{tab:baseline_comparison}
\small
\renewcommand{\arraystretch}{1.25}
\begin{tabularx}{\textwidth}{@{} l l X X @{}}
\toprule
\textbf{Classifier} & \textbf{Learning Method} & \textbf{Native Input} & \textbf{Evaluation Role} \\
\midrule

\textsf{CUMUL} & SVM with hand-crafted features
  & Statistical features and a sampled cumulative traffic curve
  & Sensitivity to aggregate traffic volume and cumulative transfer profiles \\
\addlinespace[0.4em]

\textsf{DF} & One-dimensional convolutional network
  & Direction-coded packet sequence
  & Sensitivity to local sequence patterns and positional alignment \\
\addlinespace[0.4em]

\textsf{Tik-Tok} & Deep sequence model with timing
  & Packet length, direction, and inter-arrival time (IAT)
  & Sensitivity to joint sequence and timing changes \\
\addlinespace[0.4em]

\textsf{ET-BERT} & Pre-trained Transformer
  & Anonymized datagram bytes (first five packets; 128 bytes per packet)
  & Transfer of contextualized datagram representations \\
\addlinespace[0.4em]

\textsf{YaTC} & Pre-trained masked autoencoder
  & Anonymized multi-level flow representation (first five packets; 80-byte header + 240-byte payload per packet)
  & Transfer of multi-level flow representations \\

\bottomrule
\end{tabularx}
\end{table*}

The $0\%$ and $100\%$ settings correspond to the Non-PQC and Hybrid-PQC test sets, respectively, while the four intermediate values define the Mixed test sets. For each website, samples are selected according to $\alpha$ and $1-\alpha$. Applying the same value of $\alpha$ to every website keeps the cryptographic composition uniform across all 195 labels and prevents class-dependent deployment skew from confounding classifier performance. The label space and visit-level partitioning remain fixed across the six settings, isolating the effect of changes in the test-domain composition. These test sets support the deployment-ratio and feature-control experiments in RQ3 and RQ4.

%% file: Tex/5-Evaluation.tex
\section{Evaluation}
\label{sec:evalu}

Building on the PQC-aware encrypted traffic benchmark developed in Section~\ref{sec:dataset}, this section evaluates the reliability of representative encrypted traffic classifiers under PQC-induced protocol drift across matched-domain, cross-domain, deployment-ratio, and feature-controlled settings.

\subsection{Threat Model and Evaluation Scope}

We consider a passive network adversary positioned near the client. The adversary observes the encrypted packet trace associated with a complete, isolated website visit over HTTPS, but cannot decrypt, modify, or inject traffic. Its objective is to infer the visited domain from the shared set of 195 Tranco websites under a closed-world setting. Holding the label space and browsing procedure fixed isolates the effect of the transition from Non-PQC to Hybrid-PQC key establishment on classifier reliability.

To compare the five representative classifiers without altering their native assumptions, we preserve each model's feature representation and preprocessing workflow. Inputs are restricted to information available in passively observed packets. Out-of-band DNS traffic and decrypted information from the TLS handshake or application layer, including certificate contents, HTTP Host headers, and URLs, are excluded from every classifier. The scope further excludes Tor and VPN tunneling, active attacks such as probing or traffic injection, and open-world evaluation. The experiments, therefore, assess classifier reliability under controlled PQC-induced protocol drift rather than attack effectiveness in unconstrained Internet deployments.

\subsection{Experimental Setup}

\subsubsection{Evaluated Classifiers}

To examine PQC-induced protocol drift across distinct observation surfaces and learning paradigms, we select five representative classifiers for encrypted traffic. Table~\ref{tab:baseline_comparison} organizes them into three groups consistent with the taxonomy in Section~\ref{sec:back}: an aggregate statistical method, two packet-sequence models, and two pre-trained byte-level models. This grouping reflects both the information available to each classifier and the strategy used to construct or learn its representation. Together, the models cover the packet-length, direction, timing, and byte-level dimensions, $L$, $D$, $T$, and $P$, introduced in Section~\ref{sec:analy}, while preserving their distinct native inputs.

\textsf{CUMUL} combines summary statistics with a sampled cumulative traffic curve, providing an aggregate view of transferred volume and its progression throughout a trace. It is therefore sensitive to changes in total traffic volume and cumulative transfer profiles. \textsf{DF} applies one-dimensional convolutions to direction-coded packet sequences and learns local patterns from their ordering and positional alignment. These patterns may shift if larger Hybrid-PQC key shares alter TCP segmentation and packet positions. \textsf{Tik-Tok} incorporates packet timing alongside length and direction, enabling the joint modeling of packet sequences and inter-arrival time (IAT). It provides a timing-aware view of changes arising from serialization, segmentation, and implementation-dependent processing. By comparison, \textsf{ET-BERT} pre-trains a Transformer over contextualized datagram representations. \textsf{YaTC} uses masked autoencoding to learn a multi-level flow representation. Their anonymized byte-oriented inputs represent $P$ and may be sensitive to shifted byte offsets and fixed observation windows. The five-model set therefore compares explicit side-channel representations with reusable byte-level representations across the Non-PQC and Hybrid-PQC domains, without presuming that pre-training provides cross-domain robustness.

\subsubsection{Evaluation Controls}

All five classifiers are evaluated on the same closed-world HTTPS website fingerprinting task over a shared label set of 195 websites. For each website and cryptographic domain, the evaluation draws 120 complete visits collected in Section~\ref{sec:dataset}. These visits are partitioned at the granularity of complete website visits, each of which may contain multiple TCP streams. All models use identical partition assignments, with 80, 10, and 10 ratio per website allocated to the training, validation, and test sets, respectively. 

Because the classifiers expose different input modalities by design, we preserve each model's native feature extraction and preprocessing instead of imposing a common representation. For RQ1--RQ3, \textsf{CUMUL}, \textsf{DF}, and \textsf{Tik-Tok} retain their native statistical or packet-sequence inputs. \textsf{ET-BERT} and \textsf{YaTC} retain their model-specific sanitization, tokenization, truncation, and representation construction. RQ4 deliberately varies only the input representation of \textsf{DF} while holding its architecture and training procedure fixed. Given the remaining input heterogeneity, we compare each classifier primarily against its own matched-domain baseline and avoid attributing absolute cross-model rankings solely to model architecture. We report accuracy, precision, recall, and F1 score for every evaluation setting.

\subsubsection{Research Questions}

To structure the evaluation of classifier reliability under PQC-induced protocol drift, we formulate four research questions.

\begin{tcolorbox}[
    colback=boxbg,
    colframe=boxborder,
    coltitle=black,
    colbacktitle=titlebg,
    boxrule=0.8pt,
    arc=8pt,
    title={\bfseries\sffamily Research Questions}
]
\begin{enumerate}[
    leftmargin=*,
    label=\textbf{RQ\arabic*:},
    itemsep=4pt
]
\item \textbf{(Matched-Domain Learnability)} How does Hybrid-PQC key establishment affect website fingerprinting performance when training and testing use the same cryptographic domain?

\item \textbf{(Bidirectional Cross-Domain Transfer)} How well do classifiers transfer between Non-PQC and Hybrid-PQC traffic domains?

\item \textbf{(Deployment-Ratio Sensitivity)} How does classifier performance change as the proportion of Hybrid-PQC traffic increases across the deployment-ratio settings?

\item \textbf{(Side-Channel Feature Contributions)} How do packet length, direction, and timing, individually and jointly, affect website fingerprinting performance under PQC-induced protocol drift?
\end{enumerate}
\end{tcolorbox}

\subsection{RQ1: Matched-Domain Learnability}

RQ1 examines whether website identity remains learnable after Hybrid-PQC key establishment when training and testing are matched within the same cryptographic domain. We train, validate, and test all five classifiers separately on the \textsf{Non-PQC Website Dataset} and the \textsf{PQC Website Dataset}. Both settings retain the same 195 website labels, visit-level partitions, and model configurations. The \textsf{PQC Website Dataset} contains Hybrid-PQC traffic negotiated with \texttt{\detokenize{X25519MLKEM768}}. This matched-domain design tests whether each domain supports website fingerprinting without introducing the distribution mismatch examined in RQ2. Table~\ref{tab:rq1_baselines_result} reports the results. \textbf{N-N} denotes training or fine-tuning and testing on the \textsf{Non-PQC Website Dataset}. \textbf{P-P} denotes the corresponding procedure on the \textsf{PQC Website Dataset}.

\begin{table}[htbp]
  \centering
  \setlength{\tabcolsep}{2pt} 
  \caption{Matched-domain performance of the evaluated classifiers under the N-N and P-P settings.}
  \label{tab:rq1_baselines_result}
  \begin{tabular}{lcccccccc}
    \toprule
    \multirow{2}{*}{\textbf{Classifier}} & \multicolumn{2}{c}{\textbf{Accuracy}} & \multicolumn{2}{c}{\textbf{Precision}} & \multicolumn{2}{c}{\textbf{Recall}} & \multicolumn{2}{c}{\textbf{F1 score}} \\
    \cmidrule(lr){2-3} \cmidrule(lr){4-5} \cmidrule(lr){6-7} \cmidrule(lr){8-9}
     & \textbf{N-N} & \textbf{P-P} & \textbf{N-N} & \textbf{P-P} & \textbf{N-N} & \textbf{P-P} & \textbf{N-N} & \textbf{P-P} \\
    \midrule
    CUMUL  & 0.7731 & 0.7656 & 0.7925 & 0.7731 & 0.7731 & 0.7662 & 0.7598 & 0.7538 \\
    DF     & 0.5855 & 0.5900 & 0.5792 & 0.5942 & 0.5855 & 0.5904 & 0.5500 & 0.5615 \\
    Tik-Tok & 0.5679 & 0.5737 & 0.5640  & 0.5594 & 0.5679 & 0.5739 & 0.5319 & 0.5397 \\
    ET-BERT & 0.9316 & 0.9332 & 0.9059 & 0.9076 & 0.9316 & 0.9333 & 0.9138 & 0.9157 \\
    YaTC   & 1.0000 & 0.9991 & 1.0000  & 0.9993 & 1.0000 & 0.9991 & 1.0000 & 0.9991 \\
    \bottomrule
  \end{tabular}

\end{table}

Table~\ref{tab:rq1_baselines_result} shows that each classifier attains similar N-N and P-P performance, although absolute performance differs substantially across classifiers. The aggregate-statistical \textsf{CUMUL} classifier records accuracies of $0.7731$ and $0.7656$, with F1 scores of $0.7598$ and $0.7538$. The packet-sequence models likewise show only small differences between the two domains. \textsf{DF} achieves accuracies of $0.5855$ and $0.5900$, with F1 scores of $0.5500$ and $0.5615$, under N-N and P-P, respectively. The corresponding values for \textsf{Tik-Tok} are $0.5679$ and $0.5737$ for accuracy and $0.5319$ and $0.5397$ for F1 score. Within their native preprocessing pipelines, the pre-trained byte-level models attain the highest matched-domain scores. \textsf{ET-BERT} records accuracies of $0.9316$ and $0.9332$ and F1 scores of $0.9138$ and $0.9157$. \textsf{YaTC} remains near perfect, with accuracy decreasing only from $1.0000$ to $0.9991$ and F1 score from $1.0000$ to $0.9991$. Precision and recall follow the same overall pattern. Across all four metrics, the N-N and P-P results remain close for every classifier. These matched-domain results provide the baselines for the cross-domain transfer analysis in RQ2.

\begin{tcolorbox}[
    colback=boxbg,
    colframe=boxborder,
    boxrule=0.8pt,
    arc=8pt,
]

\textbf{Answer to RQ1:} Hybrid-PQC key establishment does not produce a consistent loss of matched-domain learnability in our evaluation. All five classifiers achieve comparable N-N and P-P performance when trained and tested within a single cryptographic domain, indicating that Hybrid-PQC traffic retains learnable website fingerprints.

\end{tcolorbox}

\subsection{RQ2: Bidirectional Cross-Domain Transfer}

RQ2 evaluates whether classifiers trained in one cryptographic domain transfer to the other without retraining. The \textbf{N-P} setting models an adversary whose classifier was trained on Non-PQC traffic but is applied after Hybrid-PQC deployment. For each of the 195 websites, every classifier is trained on 80 visits and tuned on 10 validation visits from the \textsf{Non-PQC Website Dataset}. The resulting model is then evaluated on 10 held-out visits from the \textsf{PQC Website Dataset}.

The \textbf{P-N} setting reverses the source and target domains to assess whether cross-domain transfer depends on direction.
Each classifier is trained on 80 visits and tuned on 10 validation visits per website from the \textsf{PQC Website Dataset}. It is then evaluated on 10 held-out visits from the \textsf{Non-PQC Website Dataset}. Both settings use the same website labels, visit-level partitions, and model configurations as RQ1. Table~\ref{tab:rq2_baselines_result} reports the results. \textbf{N-P} denotes training or fine-tuning on Non-PQC traffic and testing on Hybrid-PQC traffic. \textbf{P-N} denotes the reverse direction. This bidirectional design distinguishes forward transfer to Hybrid-PQC traffic from backward transfer to Non-PQC traffic.

\begin{table}[htbp]
  \centering
  \setlength{\tabcolsep}{2pt} 
  \caption{Bidirectional cross-domain performance of the evaluated classifiers under the N-P and P-N settings.}
  \label{tab:rq2_baselines_result}
  \begin{tabular}{lcccccccc}
    \toprule
    \multirow{2}{*}{\textbf{Classifier}} & \multicolumn{2}{c}{\textbf{Accuracy}} & \multicolumn{2}{c}{\textbf{Precision}} & \multicolumn{2}{c}{\textbf{Recall}} & \multicolumn{2}{c}{\textbf{F1 score}} \\
    \cmidrule(lr){2-3} \cmidrule(lr){4-5} \cmidrule(lr){6-7} \cmidrule(lr){8-9}
     & \textbf{N-P} & \textbf{P-N} & \textbf{N-P} & \textbf{P-N} & \textbf{N-P} & \textbf{P-N} & \textbf{N-P} & \textbf{P-N} \\
    \midrule
    CUMUL  & 0.1919 & 0.2778 & 0.2135 & 0.2559 & 0.1928 & 0.2778 & 0.1765 & 0.2394 \\
    DF     & 0.3728 & 0.4923 & 0.3678 & 0.4800 & 0.3742 & 0.4923 & 0.3487 & 0.4609 \\
    Tik-Tok & 0.3873 & 0.4256 & 0.3971 & 0.4321 & 0.3880 & 0.4256 & 0.3627 & 0.4079 \\
    ET-BERT & 0.6255 & 0.6043 & 0.5208 & 0.5073 & 0.6256 & 0.6043 & 0.5476 & 0.5291 \\
    YaTC   & 0.3582 & 0.9709 & 0.3225 & 0.9614 & 0.3585 & 0.9709 & 0.2835 & 0.9634 \\
    \bottomrule
  \end{tabular}
\end{table}

Relative to the matched-domain baselines in Table~\ref{tab:rq1_baselines_result}, Table~\ref{tab:rq2_baselines_result} shows lower cross-domain performance for every classifier. The magnitude of degradation, however, varies by classifier and transfer direction. Under N-P, \textsf{CUMUL}, \textsf{DF}, and \textsf{Tik-Tok} achieve accuracies of $0.1919$, $0.3728$, and $0.3873$, respectively. These values are below their N-N accuracies of $0.7731$, $0.5855$, and $0.5679$. \textsf{YaTC} declines from an N-N accuracy of $1.0000$ to $0.3582$, with an N-P F1 score of $0.2835$. \textsf{ET-BERT} retains the highest N-P accuracy at $0.6255$ but remains below its N-N accuracy of $0.9316$, and its N-P F1 score is $0.5476$. Strong matched-domain performance therefore does not predict transfer from Non-PQC to Hybrid-PQC traffic.

The P-N results reveal substantial directional asymmetry. \textsf{YaTC} retains an accuracy of $0.9709$ and an F1 score of $0.9634$, close to its P-P accuracy and F1 score of $0.9991$. By contrast, \textsf{ET-BERT} records an accuracy of $0.6043$ and an F1 score of $0.5291$. \textsf{CUMUL}, \textsf{DF}, and \textsf{Tik-Tok} remain below $0.5000$ accuracy at $0.2778$, $0.4923$, and $0.4256$, respectively. Four classifiers perform better in P-N than in N-P, whereas \textsf{ET-BERT} is the exception. Thus, only \textsf{YaTC} approaches matched-domain performance in the reverse direction. The experiment establishes asymmetric transfer, but it does not by itself identify the mechanism underlying \textsf{YaTC}'s P-N behavior.


\begin{tcolorbox}[
    colback=boxbg,
    colframe=boxborder,
    boxrule=0.8pt,
    arc=8pt,
]

\textbf{Answer to RQ2:} Cross-domain transfer is generally weaker than matched-domain performance and strongly depends on the direction. All five classifiers degrade outside their training domain. Only \textsf{YaTC} approaches its matched-domain baseline in P-N, while degrading substantially in N-P. High matched-domain accuracy therefore does not ensure reliability under PQC-induced protocol drift. Models should be re-evaluated and, where necessary, retrained or adapted when the cryptographic domain changes.

\end{tcolorbox}

\subsection{RQ3: Deployment-Ratio Sensitivity}

RQ3 examines how classifiers trained or fine-tuned exclusively on Non-PQC traffic behave as the proportion of Hybrid-PQC traffic increases in the test distribution. Each of the five classifiers uses the training and validation partitions of the \textsf{Non-PQC Website Dataset} and is then evaluated, without retraining or further fine-tuning, under the six deployment-ratio settings defined in Section~\ref{sec:dataset}. Their Hybrid-PQC proportions are $\alpha \in \{0\%, 20\%, 40\%, 60\%, 80\%, 100\%\}$. All settings retain the same 195 website labels and visit-level partitions. Here, \textbf{N-N} and \textbf{N-P} denote testing at $\alpha=0\%$ and $\alpha=100\%$, respectively. The four Mixed test settings, \textbf{N-M2}, \textbf{N-M4}, \textbf{N-M6}, and \textbf{N-M8}, correspond to $\alpha=20\%$, $40\%$, $60\%$, and $80\%$. This design isolates sensitivity to deployment composition while keeping the training domain fixed.

\begin{figure}[ht]
    \centering
    \includegraphics[width=\columnwidth]{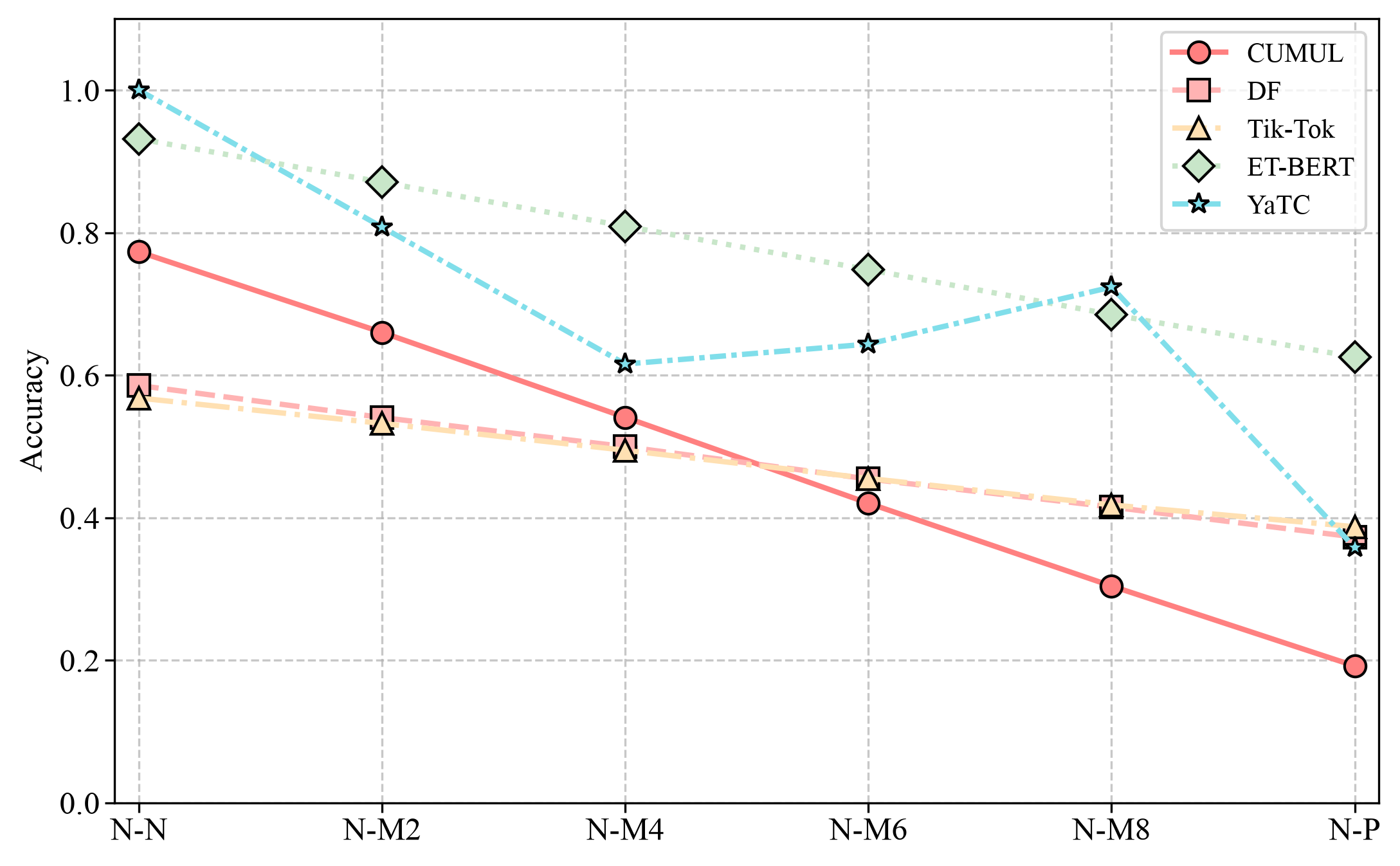}
    \caption{Accuracy trajectories of the five evaluated classifiers trained exclusively on Non-PQC traffic across six test settings as the Hybrid-PQC proportion increases from 0\% to 100\%.}
    \label{pic_rq3_histogram}
\end{figure}

Fig.~\ref{pic_rq3_histogram} shows that every classifier has lower accuracy under \textbf{N-P} than under \textbf{N-N}, although their intermediate trajectories differ. \textsf{CUMUL} declines monotonically from $0.7731$ to $0.1919$. The packet-sequence models follow shallower but similarly monotonic declines: \textsf{DF} decreases from $0.5855$ to $0.3728$, while \textsf{Tik-Tok} decreases from $0.5679$ to $0.3873$. These three classifiers therefore deteriorate progressively as the test distribution moves away from the Non-PQC training domain.

The two pre-trained byte-oriented models exhibit different responses to the changing traffic mixture:
\begin{itemize}
    \item \textsf{YaTC} achieves an accuracy of $1.0000$ under \textbf{N-N}, drops to $0.6166$ at \textbf{N-M4}, and partially recovers to $0.7252$ at \textbf{N-M8}. Its accuracy then falls to $0.3582$ under \textbf{N-P}. This non-monotonic trajectory shows that an intermediate traffic mixture does not necessarily produce intermediate performance.
    \item \textsf{ET-BERT} declines monotonically from $0.9316$ under \textbf{N-N} to $0.6255$ under \textbf{N-P}. It records the highest accuracy at \textbf{N-M2}, \textbf{N-M4}, \textbf{N-M6}, and \textbf{N-P}, whereas \textsf{YaTC} exceeds it at \textbf{N-M8}. \textsf{ET-BERT} thus exhibits the most consistent performance across the evaluated ratios, although its endpoint remains below its Non-PQC baseline.
\end{itemize}

\begin{tcolorbox}[
    colback=boxbg,
    colframe=boxborder,
    boxrule=0.8pt,
    arc=8pt,
]

\textbf{Answer to RQ3:} Across all five classifiers, increasing the Hybrid-PQC proportion generally reduces accuracy relative to the Non-PQC baseline, although the trajectories differ. \textsf{CUMUL}, \textsf{DF}, \textsf{Tik-Tok}, and \textsf{ET-BERT} decline monotonically, whereas \textsf{YaTC} changes non-monotonically. No classifier preserves its Non-PQC baseline across the full deployment range.

\end{tcolorbox}

\subsection{RQ4: Side-Channel Feature Contributions}

RQ4 isolates the individual and joint contributions of packet direction, packet length, and timing under PQC-induced protocol drift. We use \textsf{DF} as the controlled architecture and vary only its input representation. The \textbf{D} configuration retains the native direction-coded input of \textsf{DF} and serves as the baseline. We compare it with four additional configurations: (1)~\textbf{L}, using only absolute packet lengths; (2)~\textbf{T}, using only normalized inter-arrival times (IATs); (3)~\textbf{DL}, combining aligned direction and length sequences; and (4)~\textbf{DLT}, combining direction, length, and IAT. All five configurations use the same model architecture, training procedure, Non-PQC training and validation partitions, and six deployment-ratio settings defined in Section~\ref{sec:dataset} and used in RQ3. This controlled design attributes differences in classification performance to input composition rather than model architecture or dataset partitioning.

\begin{figure}[ht]
    \centering
    \includegraphics[width=\columnwidth]{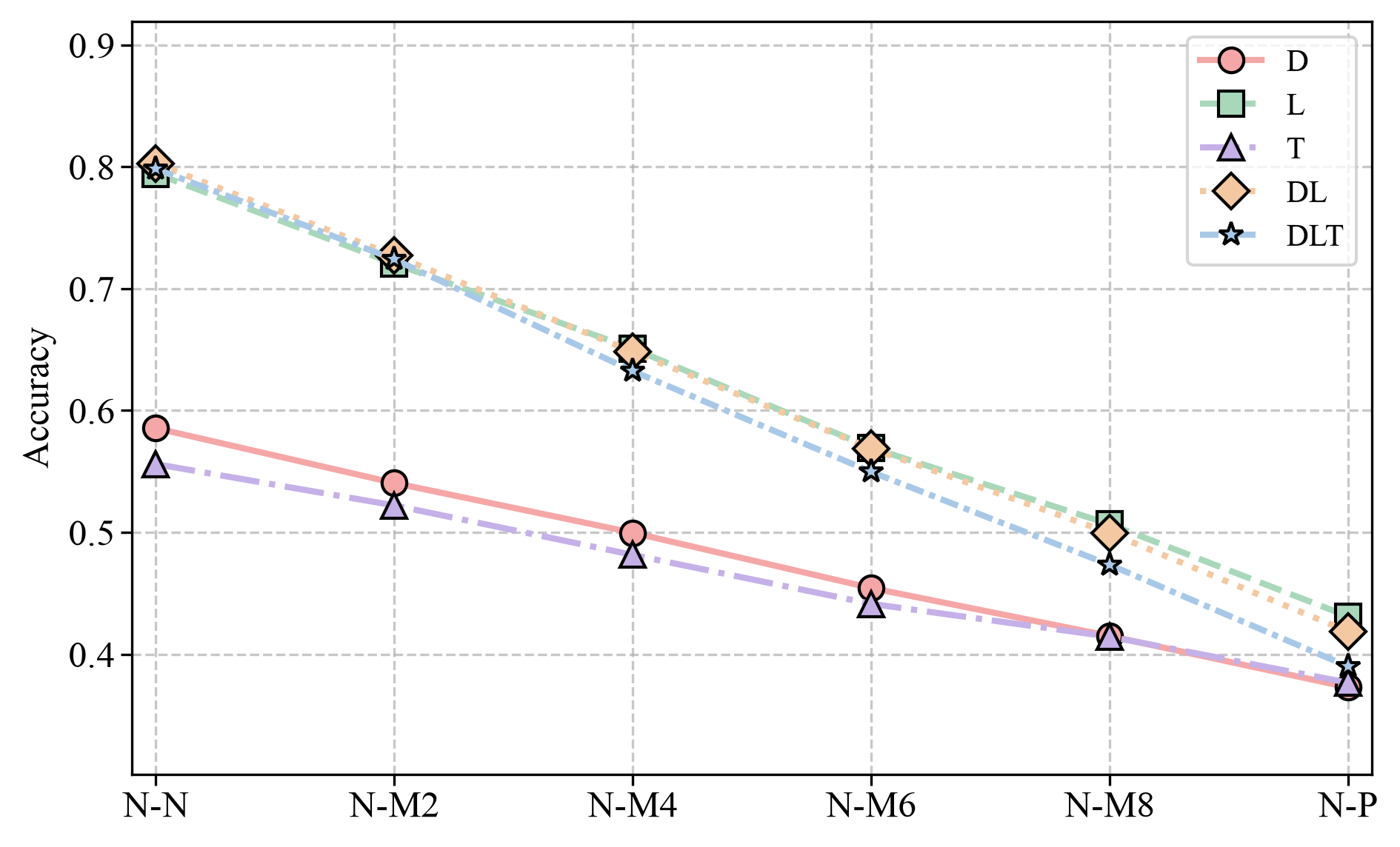}
    \caption{Accuracy trajectories of \textsf{DF} with five individual and combined side-channel representations (\textbf{D}, \textbf{L}, \textbf{T}, \textbf{DL}, and \textbf{DLT}) as the Hybrid-PQC proportion increases from 0\% to 100\%.}
    \label{pic_rq4_line}
\end{figure}

\begin{figure*}[ht]
    \centering
    \includegraphics[width=\textwidth]{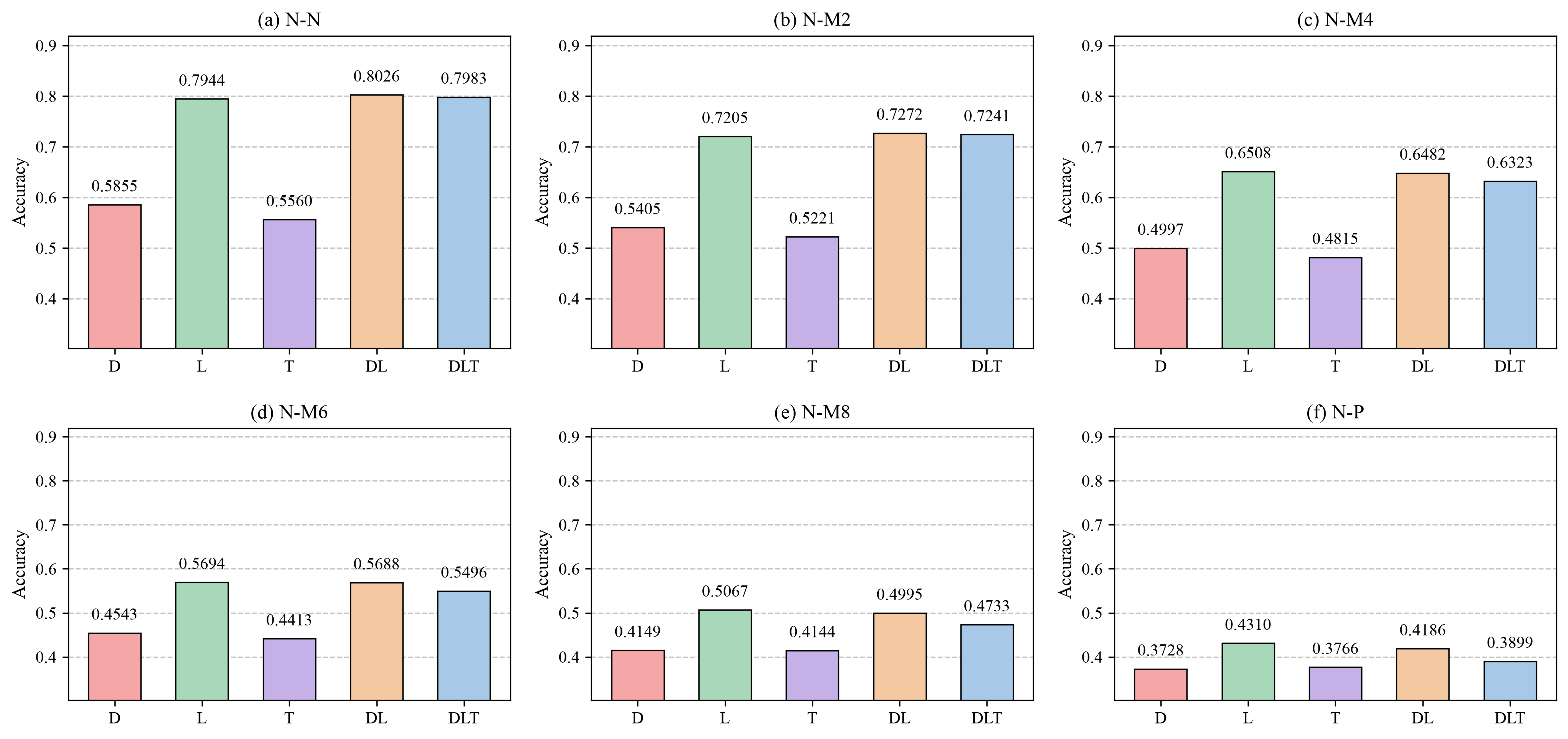}
    \caption{Per-setting accuracy of \textsf{DF} with five side-channel representations across six Hybrid-PQC deployment ratios, covering the Non-PQC, Mixed, and Hybrid-PQC test sets.}
    \label{pic_rq4_histogram}
\end{figure*}

Fig.~\ref{pic_rq4_line} shows a monotonic decline in accuracy from \textbf{N-N} to \textbf{N-P} for all five input representations. Among the single-channel configurations, \textbf{D} decreases from $0.5855$ to $0.3728$, \textbf{L} from $0.7944$ to $0.4310$, and \textbf{T} from $0.5560$ to $0.3766$. The combined representations follow the same pattern: \textbf{DL} decreases from $0.8026$ to $0.4186$, while \textbf{DLT} decreases from $0.7983$ to $0.3899$. Thus, changing the feature composition alters the attainable accuracy but does not prevent degradation as the test distribution moves away from the Non-PQC training domain.

The per-setting comparisons in Fig.~\ref{pic_rq4_histogram} identify packet length as the strongest individual feature throughout the deployment range. Under \textbf{N-N}, \textbf{L} achieves an accuracy of $0.7944$, compared with $0.5855$ for \textbf{D} and $0.5560$ for \textbf{T}. This ordering persists across all four Mixed test sets. Under \textbf{N-P}, \textbf{L} retains an accuracy of $0.4310$, whereas \textbf{D} and \textbf{T} achieve $0.3728$ and $0.3766$, respectively. However, the advantage of \textbf{L} over \textbf{D} narrows from $0.2089$ under \textbf{N-N} to $0.0582$ under \textbf{N-P}; its advantage over \textbf{T} similarly narrows from $0.2384$ to $0.0544$. Packet length therefore remains the most discriminative individual feature, although PQC-induced protocol drift progressively erodes its advantage.

Feature combination provides only marginal gains when the Hybrid-PQC proportion is low. Under \textbf{N-N}, \textbf{DL} and \textbf{DLT} achieve accuracies of $0.8026$ and $0.7983$, only slightly above the $0.7944$ achieved by \textbf{L}. The corresponding values under \textbf{N-M2} are $0.7272$, $0.7241$, and $0.7205$. From \textbf{N-M4} onward, length alone outperforms both combined representations. Under \textbf{N-P}, \textbf{L} achieves $0.4310$, compared with $0.4186$ for \textbf{DL} and $0.3899$ for \textbf{DLT}. Moreover, \textbf{DLT} never surpasses \textbf{DL}, indicating that adding timing to the direction--length representation provides no benefit in any evaluated setting. Neither direction nor timing compensates for the loss associated with PQC-induced protocol drift.

\begin{tcolorbox}[
    colback=boxbg,
    colframe=boxborder,
    boxrule=0.8pt,
    arc=8pt,
]

\textbf{Answer to RQ4:} Packet length provides the strongest individual side-channel signal across all deployment ratios, although its advantage diminishes as the Hybrid-PQC proportion increases. Direction offers no consistent benefit when combined with length, and adding timing to the direction--length representation does not improve accuracy. All five representations degrade under PQC-induced protocol drift, indicating that feature fusion alone does not provide cross-domain robustness.

\end{tcolorbox}

%% file: Tex/6-Discussion.tex
\section{Discussion}
\label{sec:dis}

\subsection{Beyond Hybrid Key Establishment: Post-Quantum Authentication}

Our empirical evaluation isolates PQC-induced drift at key establishment. It covers Hybrid-PQC negotiation with \texttt{\detokenize{X25519MLKEM768}}, while Pure-PQC key establishment appears only as a protocol-level comparison in Section~\ref{sec:analy}. A broader migration would also introduce post-quantum signatures and certificate chains into TLS authentication. ML-DSA and Falcon can add kilobyte-scale authentication material, altering packetization, burst boundaries, and handshake timing.

These effects are not uniform increases in flow size. Crossing TLS-record or TCP-segmentation thresholds can insert or reposition packets, producing discontinuities in classifier inputs. Post-quantum authentication may therefore reinforce or partially mask the drift introduced by key establishment, depending on certificate-chain composition and implementation behavior. A factorial evaluation that varies the KEM and authentication mechanism independently would separate these effects and reveal whether classifiers learn application-level invariants or migration-specific artifacts.

\subsection{Practical Trade-offs of LLM-Based Traffic Classifiers}

Model scale does not by itself address the failure mode identified in our evaluation. The results for \textsf{ET-BERT} and \textsf{YaTC} show that high matched-domain accuracy and cross-domain robustness can diverge substantially. An LLM-based classifier may capture longer byte-level dependencies and richer contextual relationships, but the same capacity may strengthen its reliance on protocol-specific shortcuts. The relevant question is whether these representations remain stable across cryptographic environments, not whether they improve matched-domain accuracy.

For a fair comparison, tokenization, pre-training corpora, input truncation, and adaptation data should be treated as part of the classifier rather than as implementation details. Evaluation should report bidirectional transfer, the data and computation required to recover from drift, and operational costs such as throughput, latency, and memory. In this setting, a larger model is practically meaningful only if it improves the robustness-cost trade-off, rather than merely recovering matched-domain accuracy through greater capacity.

\subsection{PQC-Induced Protocol Drift as Incidental Obfuscation}

The distinction between in-domain learnability and cross-domain degradation changes how PQC's apparent obfuscation should be interpreted. Hybrid-PQC traffic remains distinguishable after in-domain training, whereas Non-PQC-trained models degrade under the shifted distribution. Any apparent protection therefore reflects model staleness rather than removal of website-dependent information. Feature ablation further identifies packet length as the dominant discriminative signal, consistent with drift in message size and packetization. Deterministic cryptographic expansion is unlikely to provide durable obfuscation once its patterns enter an attacker's training data. Protocol migration nevertheless exposes a useful defensive principle: shifting traffic beyond the training distribution can impose an adaptation cost while perturbations remain difficult to model or relearn. Our closed-world evaluation of \texttt{\detokenize{X25519MLKEM768}} cannot establish this cost for other stacks or attack settings. Accordingly, defenses inspired by PQC-induced drift should be evaluated against adaptive traffic recollection and model retraining, not only fixed legacy classifiers.

%% file: Tex/7-Conclusion.tex
\section{Conclusion}
\label{sec:conclusion}

This work evaluates the reliability of encrypted traffic classifiers under PQC-induced protocol drift, focusing on the practically deployed Hybrid-PQC \texttt{\detokenize{X25519MLKEM768}} group in TLS~1.3. Across five representative classifiers, matched-domain experiments show that Hybrid-PQC traffic retains learnable website identity features when training and testing use the same cryptographic environment. However, models trained on Non-PQC traffic degrade when evaluated on Hybrid-PQC traffic, and their accuracy generally declines as the Hybrid-PQC proportion increases in Mixed test sets. The asymmetric transfer behavior of individual models further shows that strong in-domain performance does not predict stability across cryptographic environments. Our feature analysis identifies packet length as the dominant side channel, while direction and timing provide limited additional benefit in the evaluated configurations. Taken together, these findings characterize PQC evolution as a source of model mismatch rather than a permanent obstacle to encrypted traffic classification.

The results have two implications for evaluating encrypted traffic classifiers. First, protocol configurations should be treated as explicit evaluation dimensions rather than hidden dataset properties. Matched-domain accuracy should therefore be accompanied by cross-domain tests that expose sensitivity to protocol evolution. Second, datasets should reflect contemporary cryptographic deployments. The observed behavior may differ under other protocol stacks, cryptographic configurations, or attack settings not examined in this work. More broadly, reliable encrypted traffic classification requires evidence that representations transfer across evolving protocol environments, not merely that they perform well under matched conditions.